\documentclass{IEEEtaes}

\usepackage{color,array,amsthm}
\usepackage{graphicx}
\usepackage{amsmath}
\usepackage{tabularx} % Include this in the preamble
\usepackage{xcolor}
\usepackage{tikz}
\usepackage{lettrine}
\usepackage{textcomp}
\usepackage{stfloats}
\usepackage{lipsum}  
\usepackage{verbatim}
\usepackage{graphicx}
\usepackage{cite}
\usepackage{pdfpages}
\usepackage{subfig}

\jvol{XX}
\jnum{XX}
\jmonth{XXXXX}
\paper{1234567}
\pubyear{2022}
\doiinfo{TAES.2022.Doi Number}

\begin{document}

\title{Martian Dust Storm Detection with THz Opportunistic Integrated Sensing and Communication in the Internet of Space (IoS)}

\author{Haofan Dong}
\member{Student Member, IEEE}
\affil{University of Cambridge, Cambridge, CB3 0FA, UK} 

\author{Ozgur B. Akan}
\member{Fellow, IEEE}
\affil{University of Cambridge, Cambridge, CB3 0FA, UK, and Koç University, Istanbul, 34450, Turkey}

%% \author{FOURTH D. AUTHOR}
%% \affil{University of Colorado, Colorado, USA}

\receiveddate{Manuscript received XXXXX 00, 0000; revised XXXXX 00, 0000; accepted XXXXX 00, 0000.\\}
% This paragraph of the first footnote will contain the date on which you submitted your paper for review, which is populated by IEEE. It is IEEE style to display support information, including sponsor and financial support acknowledgment, here and not in an acknowledgment section at the end of the article. For example, ``This work was supported in part by the U.S. Department of Commerce under Grant BS123456.'' }
%% \accepteddate{XXXXX XX XXXX}
%% \publisheddate{XXXXX XX XXXX}

\corresp{Corresponding author: Ozgur B. Akan (oba21@cam.ac.uk)}

\authoraddress{
    Haofan Dong is with the Internet of Everything Group, Department of Engineering, University of Cambridge, CB3 0FA Cambridge, UK (e-mail: hdong@cam.ac.uk). 
    Ozgur B. Akan is with the Internet of Everything Group, Department of Engineering, University of Cambridge, CB3 0FA Cambridge, UK, and also with the Center for neXt-generation Communications (CXC), Department of Electrical and Electronics Engineering, Koç University, 34450 Istanbul, Turkey (e-mail: oba21@cam.ac.uk).
}

% \editor{Mentions of supplemental materials and animal/human rights statements can be included here.}
% \supplementary{Color versions of one or more of the figures in this article are available online at {http://ieeexplore.ieee.org}.}

\markboth{DONG ET AL.}{MARS DUST DETECTOR}
\maketitle

\begin{abstract}This paper presents the Mars Dust Storm Detector (MDSD), a system that leverages the THz Opportunistic Integrated Sensing and Communications (OISAC) signals between Mars surface assets (rovers and landers) to extract environmental information, particularly dust storm properties. The MDSD system utilizes the multi-parameter sensitivity of THz signal attenuation between Martian communication devices to provide rich, real-time data on storm intensity, particle characteristics, and potentially even electrification state. This approach, incorporating HITRAN spectroscopic data and Martian-specific atmospheric parameters, allows for accurate modeling and analysis. The system's ability to repurpose THz ISAC signals for environmental sensing demonstrates an efficient use of resources in the challenging Martian environment, utilizing communication infrastructure to enhance our understanding of Mars' atmospheric dynamics. The system's performance is evaluated through extensive simulations under various Node Density Factors (NDFs), comparing different interpolation algorithms for dust storm intensity mapping. Results demonstrate that linear interpolation achieves superior accuracy (correlation >0.90) at high NDFs, while nearest-neighbor and IDW algorithms maintain complete spatial coverage in sparse networks. Error analysis identifies dust particle size uncertainty as the primary contributor to estimation errors, though the system shows resilience to Martian atmospheric variations. This work extends the opportunistic use of ISAC technology to planetary exploration, contributing to both Mars atmospheric monitoring capabilities and ISAC applications in the Internet of Space (IoS).
\end{abstract}


\section{INTRODUCTION}

T{\scshape he} advancement of space technology has enabled deep space science missions, including the exploration of other celestial bodies. These missions generate substantial data volumes, necessitating advanced communication technologies for transmission to Earth within the Internet of Space (IoS) framework \cite{IoSp}. Mars, with its unique atmosphere and geology, presents a prime target for human exploration while posing significant challenges \cite{wedage2022path}. As discussed in \cite{goldstein1968communication}, the communication capability from Mars is fundamentally limited by basic phenomena and practical considerations, especially for real-time data transmission. Years of Martian missions utilizing rovers and landers have greatly enhanced our understanding of the planet's surface, atmosphere, and habitability potential. Mars experiences more frequent and intense dust storms compared to Earth, which not only pose risks to detection equipment but also offer valuable insights into Martian atmospheric dynamics \cite{doexamining, balme2006dust}. The study of these storms is crucial for comprehending the Martian environment and planning future manned missions.

In the realm of emerging communication technologies, Integrated Sensing and Communications (ISAC) has emerged as a key component of 6G systems \cite{liu2022integrated}. ISAC integrates communication capabilities with environmental sensing, offering numerous applications while optimizing hardware and spectrum utilization \cite{liu2022survey}. The Terahertz (THz) frequency range, spanning 0.3-10 THz, presents a promising solution for the demanding conditions on Mars. The superior transmission capabilities of the THz band, coupled with the compactness of associated hardware, align well with the operational requirements of Martian expeditions \cite{pawar2013terahertz}.

On Earth, the potential of ISAC for environmental sensing has already been demonstrated in communication networks that leverage signal attenuation \cite{messer2006environmental, messer2015new} and frequency-specific molecular absorption to monitor atmospheric conditions \cite{wedage2021climate}. Research conducted in Europe \cite{leijnse2007rainfall} and Israel \cite{david2009novel} has shown that Received Signal Level (RSL) data from terrestrial microwave communication links can effectively estimate spatial and temporal rainfall intensity. This technique is based on the established power law relation $A = aR^{b}$ \cite{olsen1978ar}, which links the attenuation of the microwave signal (A) to the intensity of rainfall (R) \cite{messer2006environmental}. In the context of space applications, our parallel work \cite{dong2024debrisense} has demonstrated the potential of THz ISAC techniques for detecting and classifying space debris in Low Earth Orbit (LEO) satellite networks, highlighting the versatility of this approach in space-based sensing.

The propagation characteristics of the THz band have been shown to be more favourable on Mars than on Earth due to Mars' arid atmosphere and vast terrain \cite{diao2021comparison,wedage2023comparative}. Recent work in \cite{tekbiyik2022wireless} has demonstrated that these effects are particularly pronounced during the southern hemisphere's late spring and early summer, when intense dust activity is driven by strong thermal gradients and surface winds. This presents an intriguing opportunity to exploit THz link attenuation for assessing Martian dust storm intensity. Conventional methods for monitoring Martian dust storms, such as optical instruments on rovers and high-resolution and infrared cameras on orbiters \cite{wang2015origin}, face challenges in geographic resolution, incur significant costs, and are subject to considerable uncertainties.

This paper presents the Mars Dust Storm Detector (MDSD), an innovative system that leverages THz Opportunistic Integrated Sensing and Communications (OISAC) signals between Mars surface assets (rovers, landers, and orbiters) to extract environmental information, particularly dust storm properties. The MDSD system utilizes the multi-parameter sensitivity of THz signal attenuation between Martian communication devices to provide rich, real-time data on storm intensity, particle characteristics, and potentially even electrification state. Key contributions of this work include the following.

\begin{enumerate}
\item Development of a comprehensive THz signal attenuation model for the Martian atmosphere, incorporating factors such as dust particle density, atmospheric pressure, and temperature variations.
\item Design and analysis of the MDSD system architecture, demonstrating the feasibility of repurposing THz ISAC signals for environmental sensing on Mars.
\item Comparative analysis of various interpolation algorithms for dust storm intensity mapping, considering different Node Density Factors (NDFs) in different storm seasons.
\item Performing error analysis quantifying the impact of uncertainties in dust particle properties, atmospheric conditions, and measurement processes on system performance.
\end{enumerate}

The paper is organized as follows: Section II presents the mathematical model for THz signal attenuation in the Martian atmosphere, considering both dust storm and molecular absorption effects. Section III details the MDSD system architecture and data processing methods, along with system analysis. Section IV describes the evaluation metrics and simulation configurations used to assess the system's performance. Section V presents and discusses the simulation results, comparing different interpolation algorithms and analyzing the system's sensitivity to various environmental factors. Finally, Section VI concludes the paper and outlines future research directions.

\section{Signal Attenuation Modelling for Martian Dust Storms}
% Environmental factors such as atmospheric molecules, temperature, pressure, and airborne particles play a crucial role in the propagation of wireless signals \cite{wedage2023comparative}. On Mars, these challenges are amplified by the planet's thin atmosphere and frequent, intense dust storms that can scatter and absorb electromagnetic waves. The lower atmospheric pressure and cooler temperatures on Mars, compared to Earth, further exacerbate these issues. 

% Dust activity on Mars is particularly intense during the late spring and early summer in the southern hemisphere, driven by strong winds and significant temperature variations \cite{shekh2021effect}.

% This section delves into the attenuation of THz signals under Martian surface conditions, with a special focus on the impact of dust storms. A point-to-point propagation model serves as the theoretical framework, offering a simplified yet effective basis for analysing signal behaviour on Mars.

The propagation of THz signals in the Martian atmosphere is primarily affected by dust particles and molecular absorption, with unique challenges posed by Mars' thin atmosphere and frequent dust storms \cite{wedage2023comparative}. These effects are particularly pronounced during the southern hemisphere's late spring and early summer, when intense dust activity is driven by strong thermal gradients and surface winds \cite{shekh2021effect}.

\subsection{Attenuation due to Dust Storms}

% Research on the attenuation effects of the THz and sub-THz frequency bands by dust particles in the propagation path has been extensive, aiming to elucidate the impact of dust storms on wireless communication links. Various models have been advanced to simulate these effects \cite{alozie2023review}, with the Rayleigh approximation and the Mie theory \cite{goldhirsh1982parameter, chu1979bstj} standing out for their applicability. The Rayleigh approximation proves effective for particles of spherical and elliptical shapes, yet its relevance wanes for particle sizes nearing or surpassing 10\% of the signal's wavelength. In contrast, the Mie theory, which considers Mie scattering, is better suited for instances where particle dimensions exceed wavelength, offering a more accurate determination of the effective refractive index \cite{goldhirsh2001attenuation}.

% Considering the Martian environment, where the mean diameters of the dust particles hover around 4 µm \cite{lemmon2019martian}, and given that these dimensions fall below one-tenth of the THz signal wavelength employed in our analysis, the Rayleigh approximation serves as a viable method for approximating the total extinction cross section of the Martian dust.

The interaction between THz waves and Martian dust particles can be characterized through two principal theoretical frameworks: the Rayleigh approximation and Mie theory \cite{alozie2023review}. For particles significantly smaller than the wavelength ($D < 0.1\lambda$), the Rayleigh approximation provides an efficient model for spherical and elliptical particles. Conversely, Mie theory offers more accurate modeling when particle dimensions approach or exceed the wavelength \cite{goldhirsh2001attenuation}.

The signal attenuation is characterized through the complex refractive index ($n$) and dielectric constant ($\varepsilon$):
\begin{equation} \label{eq:refractive_dielectric}
n=n^{\prime}-jn^{\prime \prime}, \quad \varepsilon=\varepsilon^{\prime}-j\varepsilon^{\prime \prime}
\end{equation}
where $\varepsilon^{\prime}$ and $\varepsilon^{\prime \prime}$ represent the real and imaginary parts of the complex permittivity.

The attenuation coefficient $A_{dust}$, expressing signal strength reduction per unit length, is defined in nepers/m and dB/m:
\begin{align} 
A_{dust} &= k \cdot n^{\prime \prime} \quad (\text{nepers/m}) \label{eq:attenuation_coefficient} \\
A_{dust} &= 8.686 \cdot k \cdot n^{\prime \prime} \quad (\text{dB/m}) \label{eq:attenuation_dbm}
\end{align}
where $k=2\pi/\lambda$ is the wave number.

For forward scattering analysis, the refractive index incorporates the scattering function $S(0)$ \cite{hulst1981light,smith1986propagation}:
\begin{align}
n &= 1-i \cdot S(0) \cdot 2\pi N \cdot k^{-3} \label{eq:particle_scattering} \\
S(0) &= i \cdot k^3\left(\frac{\varepsilon-1}{\varepsilon+2}\right)r^3+\frac{2}{3}k^6\left(\frac{\varepsilon-1}{\varepsilon+2}\right)^2r^6 \label{eq:scattering_factor}
\end{align}

Neglecting higher-order terms in (\ref{eq:scattering_factor}) and combining with (\ref{eq:particle_scattering}), we obtain:
\begin{equation} \label{eq:further_refractive}
n=1+\left(\frac{\varepsilon-1}{\varepsilon+2}\right)r^3 \cdot 2\pi N
\end{equation}

The final expression for dust-induced attenuation, considering average particle radius $\bar{r}$, is:
\begin{equation} \label{eq:final_attenuation}
A_{dust}=\frac{1.029 \times 10^6 \cdot \varepsilon^{\prime \prime}}{\left[\left(\varepsilon^{\prime}+2\right)^2+\varepsilon^{\prime \prime 2}\right] \cdot \lambda} \cdot N \cdot \bar{r}^3 \, (\text{dB/km})
\end{equation}

\subsection{ Attenuation by Molecular Absorption}

In the THz frequency range, molecular absorption plays a crucial role in signal attenuation. Unlike Earth's atmosphere where H\textsubscript{2}O and O\textsubscript{2}dominate THz absorption, the Martian atmosphere presents a unique scenario due to its high CO\textsubscript{2} concentration (95.32\%) with trace amounts of N\textsubscript{2} and other gases. Despite this high CO\textsubscript{2} concentration, the extremely low atmospheric pressure (0.6\% of Earth's) results in relatively mild molecular absorption of THz signals. This creates potentially favorable propagation conditions for THz ISAC systems on Mars for two reasons: (1) the dry atmosphere contains substantially lower levels of water vapor and oxygen, which are key absorbers of THz radiation on Earth, and (2) the low pressure broadening creates distinct transmission windows between CO\textsubscript{2} absorption lines \cite{o2019perspective,smith2008martian}.

The molecular absorption loss, quantifying the power of electromagnetic waves converted into kinetic energy by molecular vibrations, can be expressed using the Beer-Lambert law for a homogeneous medium between a transmitter and a receiver over distance $d$ at frequency $f$ \cite{5995306}:
\begin{equation}
A_{\text{abs}}(f,d) = e^{k(f) \cdot d}
\label{eq:abs_loss}
\end{equation}
where $k(f)$ is the frequency-dependent absorption coefficient.

The absorption coefficient is a sum of contributions from various gas species and their isotopes present in the Martian atmosphere:
\begin{equation}
k(f) = \sum_{g,i} k_g^i(f)
\label{eq:abs_coeff}
\end{equation}

Here, $k_g^i(f)$ represents the monochromatic absorption coefficient of the $g^{\text{th}}$ gas's $i^{\text{th}}$ isotope at frequency $f$. To accurately model this coefficient, we consider several factors:

\begin{enumerate}
    \item \textbf{Temperature-dependent line intensity}: The line intensity $S_g^i(T)$ at Martian temperature $T$ is calculated from the reference intensity $S_g^i(T_0)$ at $T_0 = 296$ K, accounting for the lower state energy $E_l$ and the partition function ratio $Q_{\text{ratio}}$ \cite{gordon2022hitran}:
    \begin{equation}
S_g^i(T) = S_g^i(T_0)  Q_{\text{ratio}}  e^{-\frac{hcE_l}{k} \left(\frac{1}{T} - \frac{1}{T_0}\right)} \frac{1 - e^{-\frac{hc\nu}{kT}}}{1 - e^{-\frac{hc\nu}{kT_0}}}
\label{eq:line_intensity}
\end{equation}
    
    \item \textbf{Isotopic abundance and atmospheric composition}: The line intensity is adjusted for the specific isotopic abundance and the mixing ratio in the Martian atmosphere \cite{mahaffy2013abundance}.
    
    \item \textbf{Line broadening}: In the low-pressure Martian environment, Doppler broadening is dominant. The Doppler half-width $a_D^i$ is given by \cite{wedage2023comparative}:
    \begin{equation}
    a_D^i = \frac{f_g^i}{c} \cdot \sqrt{\frac{2N_A \cdot k_B \cdot T \cdot \ln 2}{M^i}}
    \label{eq:doppler_width}
    \end{equation}
    where $f_g^i$ is the resonance frequency, $M^i$ is the molecular mass, $N_A$ is the Avogadro constant and $k_B$ is the Boltzmann constant.
    
    \item \textbf{Line shape}: Due to the low-pressure environment on Mars, a Gaussian profile is assumed for the line shape \cite{wedage2023comparative}:
    \begin{equation}
    F_g^i(f) = \sqrt{\frac{\ln 2}{\pi \cdot {a_D^i}^2}} \cdot e^{-\frac{(f-f_g^i)^2 \ln 2}{{a_D^i}^2}}
    \label{eq:line_shape}
    \end{equation}
\end{enumerate}

The absorption cross-section $\sigma_g^i(f)$ is then computed as the product of the temperature-dependent line intensity and the line shape function:
\begin{equation}
\sigma_g^i = S_g^i(T) \cdot F_g^i(f)
\label{eq:abs_cross_section}
\end{equation}

For the CO\textsubscript{2}-dominated Martian atmosphere, the absorption characteristics differ significantly from Earth. While CO\textsubscript{2} molecules exhibit numerous absorption lines in the THz band, the low atmospheric pressure (approximately 610 Pa) leads to narrower linewidths and reduced overall absorption compared to Earth's water vapor absorption. This unique atmospheric composition influences the selection of optimal frequency bands for the MDSD system.

Finally, the monochromatic absorption coefficient is determined by:
\begin{equation}
k_g^i(f) = \frac{p}{p_0} \cdot \frac{T_{\text{STP}}}{T} \cdot Q_g^i \cdot \sigma_g^i(f)
\label{eq:mono_abs_coeff}
\end{equation}
where $p$ and $T$ are the Martian atmospheric pressure and temperature, $p_0$ and $T_{\text{STP}}$ are standard pressure and temperature, and $Q_g^i$ is the molecular volume density, calculated using the ideal gas law:
\begin{equation}
Q_g^i = \frac{p}{R \cdot T} \cdot q_g^i \cdot N_A
\label{eq:mol_vol_density}
\end{equation}
where $q_g^i$ is the mixing ratio of isotope $i$ of gas $g$.

This approach, incorporating HITRAN spectroscopic data and Mars-specific atmospheric parameters, allows for accurate modelling of the attenuation of the THz signal in the Martian atmosphere.

\subsection{Unified Attenuation Model}

The total THz signal attenuation in the Martian environment combines dust-induced and molecular absorption effects:
\begin{equation} \label{eq:total_attenuation}
A_{\text{total}} = A_{\text{dust}}(f, N) + A_{\text{abs}}(f, T, p)
\end{equation}
where $A_{\text{dust}}$ represents dust particle attenuation and $A_{\text{abs}}$ accounts for molecular absorption, both dependent on frequency $f$, particle concentration $N$, temperature $T$, and pressure $p$.

Isolating the dust-induced component:
\begin{equation} \label{eq:dust_attenuation}
A_{\text{dust}} = A_{\text{total}} - A_{\text{abs}}
\end{equation}

Substituting the derived expressions yields:
\begin{equation} \label{eq:attenuation_equation}
A_{\text{total}} - k(f) \cdot 10^4 \log_{10}{e} = \frac{1.029 \times 10^6 \cdot \varepsilon^{\prime\prime}}{\left[(\varepsilon^{\prime}+2)^2+\varepsilon^{\prime\prime 2}\right] \cdot \lambda} \cdot N \cdot r^3
\end{equation}

To enable practical measurements, we relate particle concentration to visibility through \cite{ho2002radio}:
\begin{equation} \label{eq:particle_concentration}
N = \frac{5.5 \times 10^{-4}}{r^2 \cdot V}
\end{equation}

This yields the comprehensive attenuation-visibility relationship:
\begin{equation} \label{eq:final_V_equation}
V = \frac{566 \cdot r \cdot \varepsilon^{\prime\prime}}{\left(A_{\text{total}} - k(f) \cdot 10^4 \log_{10}{e}\right) \cdot \lambda \cdot \left[(\varepsilon^{\prime}+2)^2+\varepsilon^{\prime\prime 2}\right]}
\end{equation}

Equation \eqref{eq:final_V_equation} enables dust storm characterization through THz link measurements, providing a quantitative basis for storm intensity estimation from communication performance metrics.

\section{Mars Dust Storm Detector (MDSD) System}

Building upon the networked radar concepts presented in \cite{sahin2024resource}, the MDSD system employs THz ISAC signals between Mars surface assets to detect and characterize dust storms. This section presents the system architecture and analysis framework.

\subsection{System Architecture}

\begin{figure}[t]
\centering
\includegraphics[width=1\columnwidth]{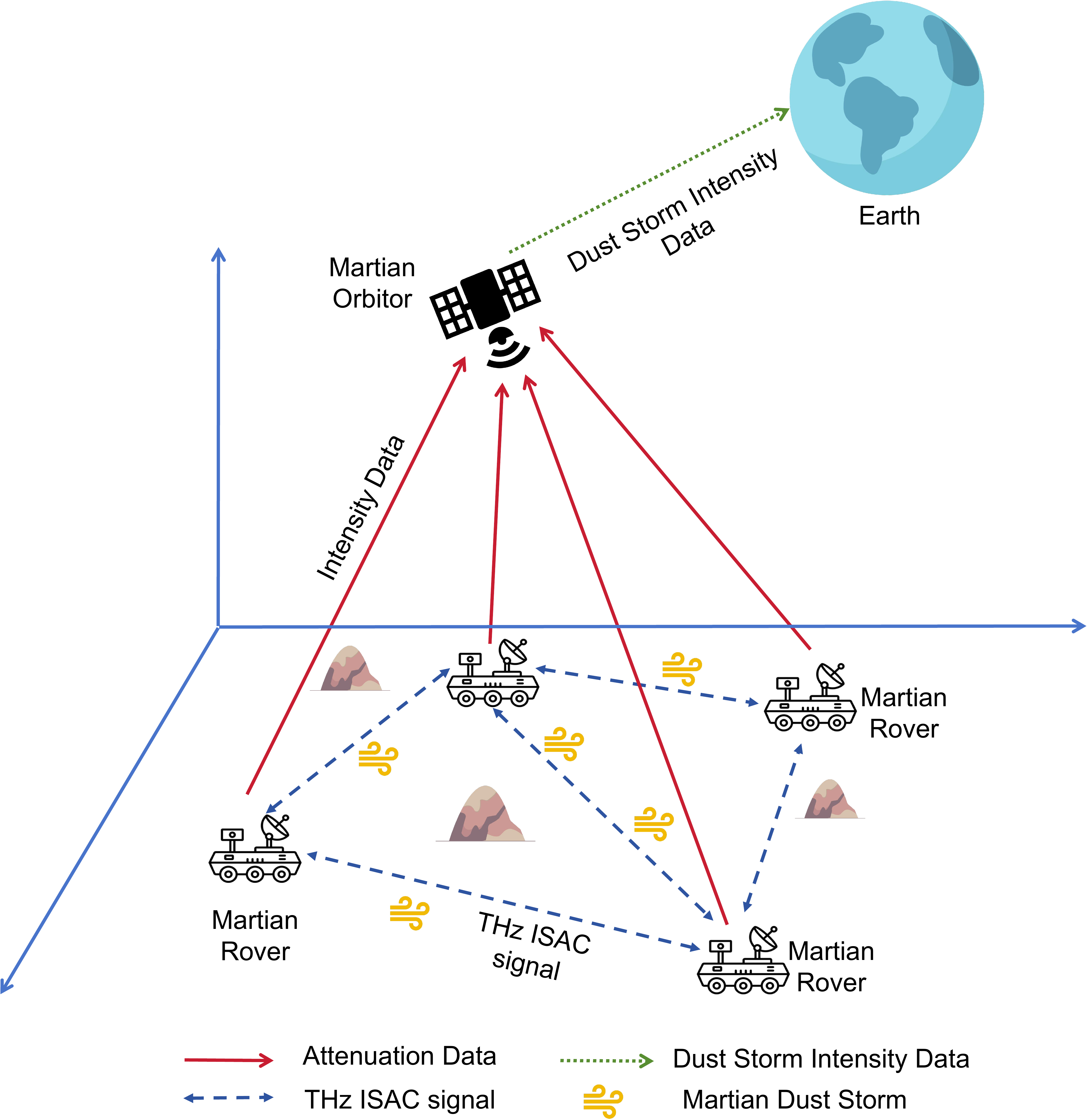}
\caption{MDSD system architecture: THz ISAC signals between Martian rovers experience dust storm attenuation. The orbiter collects and relays intensity data to Earth.}
\label{fig:system_architecture}
\end{figure}

\begin{figure*}[h]
\centering
\includegraphics[width=1\textwidth]{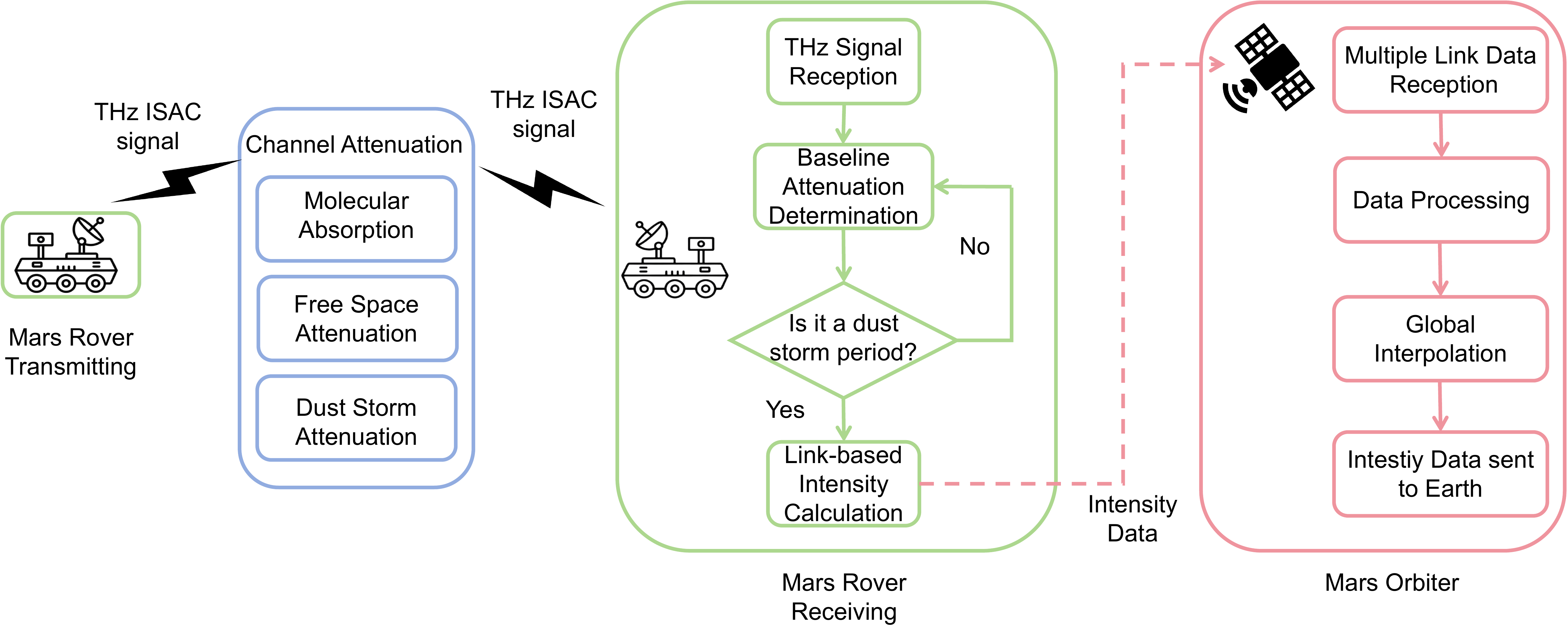}
\caption{MDSD System Logic Diagram: Data flow from THz signal transmission to dust storm intensity calculation and global interpolation.}
\label{fig:MDSD_logic}
\end{figure*}

The MDSD system utilizes distributed THz links between rovers and landers for sensing, while orbiters aggregate data for global monitoring (Fig. \ref{fig:system_architecture}). The signal processing chain, from THz transmission to storm mapping, is illustrated in Fig. \ref{fig:MDSD_logic}.

The fundamental principle relies on the relationship between signal attenuation and dust storm properties. The total attenuation comprises:

\begin{equation}
A_{\text{total}} = A_{\text{dust}}(f,N) + A_{\text{abs}}(f,T,p) + A_{\text{fs}}
\label{eq:total_attenuation}
\end{equation}

where $A_{\text{dust}}$ represents dust-induced attenuation dependent on frequency $f$ and particle density $N$, $A_{\text{abs}}$ accounts for molecular absorption varying with temperature $T$ and pressure $p$, and $A_{\text{fs}}$ denotes free space path loss.

The system operates through three key processes:
\begin{enumerate}
\item \textit{Signal Acquisition}: Rovers monitor received signal strength, establishing baseline attenuation during clear conditions:
\begin{equation}
A_0(h) = \text{median}\{A_M(t) | t \in S_0, \text{hour}(t) = h\}
\label{eq:baseline}
\end{equation}
where $A_M(t)$ is measured attenuation at time $t$, and $S_0$ denotes dust-free periods.

\item \textit{Dust Detection}: Multi-link correlation analysis identifies dust events through:
\begin{equation}
\bar{\rho} > \rho_{\text{threshold}} \text{ AND } \overline{\Delta A} < -\alpha \text{ dB/km}
\label{eq:detection}
\end{equation}
with empirically determined thresholds $\rho_{\text{threshold}}$ and $\alpha$.

\item \textit{Storm Mapping}: Interpolation techniques generate continuous intensity maps:
\begin{equation}
\theta(x,y) = \frac{\sum_{i,j} (W_{ij} + z\sigma_i^2)^{-1} r_{ij}}{\sum_{i,j} (W_{ij} + z\sigma_i^2)^{-1}}
\label{eq:mapping}
\end{equation}
incorporating measurement uncertainty $\sigma_i^2$ in the weighting scheme.
\end{enumerate}

This architecture enables comprehensive dust storm monitoring by repurposing communication signal, providing real-time, wide-area coverage of Martian atmospheric dynamics.

\subsection{THz Propagation Model}

The dust-induced attenuation, key to storm sensing, is characterized by: 

The dominance of CO\textsubscript{2} in the Martian atmosphere presents both advantages and challenges for THz propagation. The CO\textsubscript{2} absorption peaks are well-defined and relatively narrow due to low pressure broadening, creating exploitable transmission windows between absorption lines. Additionally, the temporal stability of CO\textsubscript{2} concentration, compared to Earth's variable water vapor, enables more reliable baseline calibration for dust storm detection.

The key component for dust storm sensing is the dust-induced attenuation:

\begin{equation}
A_{\text{dust}} = \frac{1.029 \times 10^6 \cdot \varepsilon''}{\left[(\varepsilon' + 2)^2 + \varepsilon''^2\right] \cdot \lambda} \cdot N \cdot \bar{r}^3
\label{eq:dust_attenuation_sensing}
\end{equation}

where the parameters reflect Martian dust properties: $\bar{r} = 4 \times 10^{-6}$ m, $\varepsilon' = 1.55$, and $\varepsilon'' = 6.3$ \cite{goldhirsh1982parameter}.

\begin{figure}[t]
\centering
\includegraphics[width=\columnwidth]{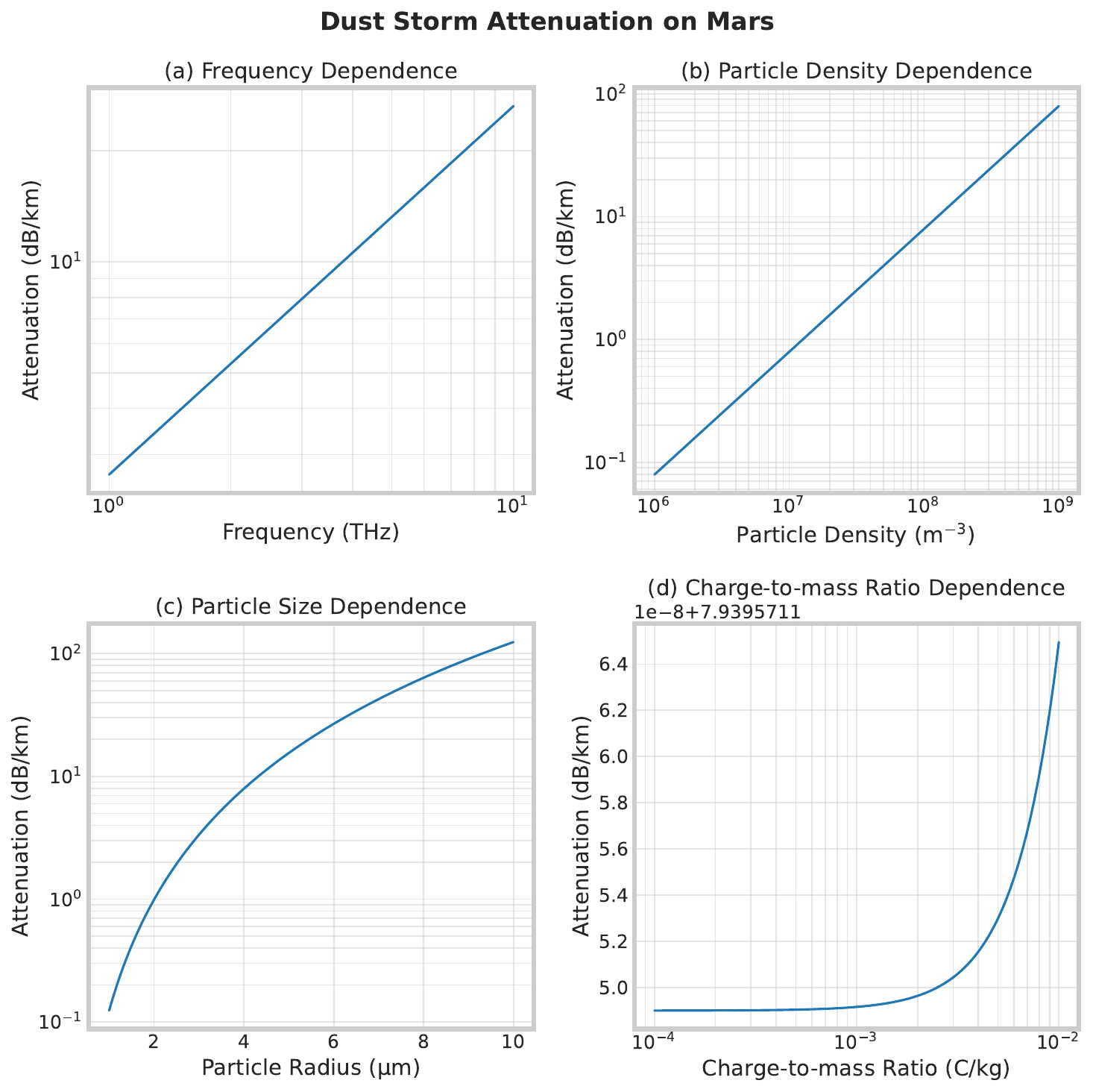}
\caption{Dust storm parameter sensitivity analysis showing attenuation dependence on: (a) frequency, (b) particle density, (c) particle size, and (d) charge-to-mass ratio.}
\label{fig:dust_sensitivity}
\end{figure}

The system's sensitivity to dust storm parameters is characterized through four key relationships, as shown in Fig. \ref{fig:dust_sensitivity}:

\begin{enumerate}
\item Frequency response: Linear increase in attenuation with frequency, offering enhanced sensitivity at higher frequencies
\item Particle density correlation: Direct relationship enabling dust concentration estimation
\item Size distribution effects: Non-linear response to particle radius variations
\item Charge sensitivity: Sharp increase in attenuation beyond $10^{-3}$ C/kg charge-to-mass ratio
\end{enumerate}

For practical implementation, the model incorporates Martian atmospheric variations:
\begin{itemize}
\item Temperature range: 180-280 K
\item Pressure variation: 400-700 Pa
\item Primary atmospheric constituents: CO$_2$ (95.32\%), N$_2$ (2.7\%), Ar (1.6\%)
\end{itemize}

The simulation results demonstrate that despite the high CO$_2$ concentration in Mars' atmosphere, its impact on THz signal propagation is manageable. This is primarily due to three factors: (1) the extremely low atmospheric pressure reducing pressure broadening effects, (2) the stability of CO$_2$ concentration allowing for effective baseline calibration, and (3) the existence of multiple transmission windows between CO$_2$ absorption lines suitable for ISAC operations.

These environmental parameters are integrated into the molecular absorption calculation through the modified HITRAN database coefficients for Martian conditions, enabling accurate distinction between dust-induced and atmospheric attenuation effects.

\begin{figure*}[h!]
    \centering
    \includegraphics[width=\textwidth]{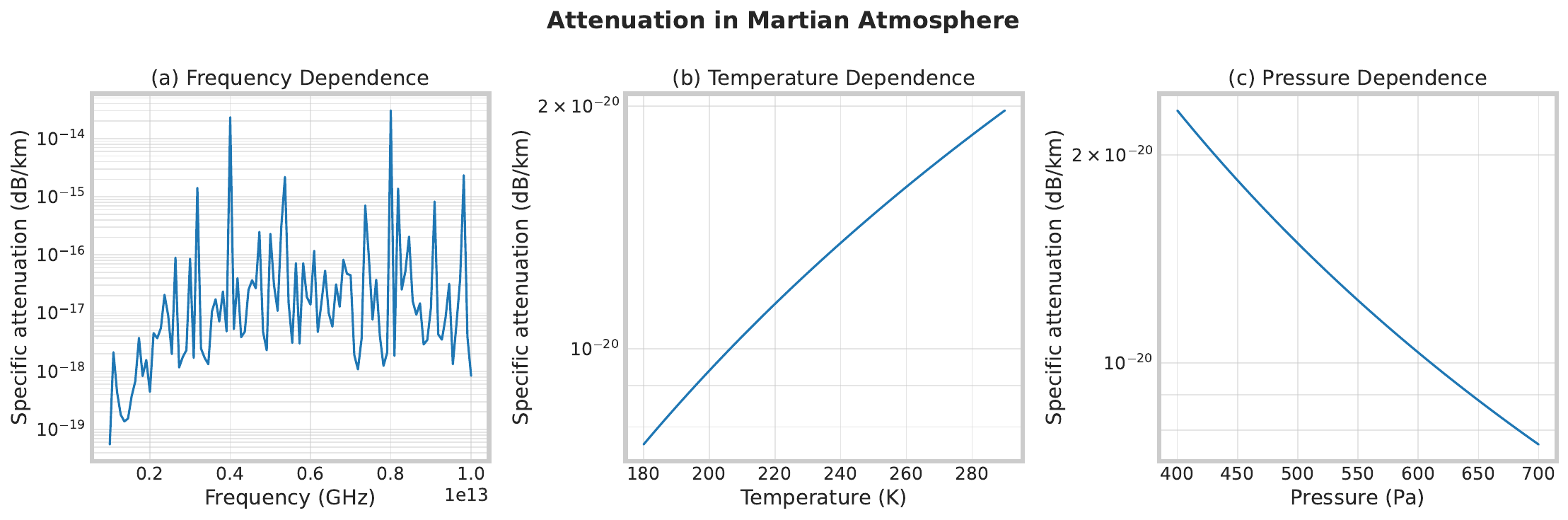}
    \caption{Attenuation in Martian Atmosphere: (a) Frequency Dependence, (b) Temperature Dependence, and (c) Pressure Dependence. These plots illustrate how specific attenuation varies with frequency, temperature, and pressure, highlighting key atmospheric influences on THz signal propagation on Mars.}
    \label{fig:martian_atmosphere_attenuation}
\end{figure*}

\subsection{Dust Storm Detection and Characterization}

The MDSD system employs a multi-stage approach for dust storm detection and characterization, incorporating correlation-based detection, intensity quantification, and error analysis.

\subsubsection{Detection Algorithm}
Dust storm events are identified through cross-correlation analysis of multiple THz links. The correlation coefficient between link pairs is computed as:

\begin{equation}
\rho_{i,j} = \frac{\text{Cov}(\Delta A_i, \Delta A_j)}{\sigma_{\Delta A_i} \sigma_{\Delta A_j}}
\label{eq:correlation}
\end{equation}

where $\Delta A_i$ represents attenuation changes in link $i$. A dust storm event is confirmed when the network-wide correlation $\bar{\rho}$ exceeds a threshold and significant attenuation is observed:

\begin{equation}
\bar{\rho} > 0.7 \text{ AND } \overline{\Delta A} < -\alpha \text{ dB/km}
\label{eq:storm_criteria}
\end{equation}

\subsubsection{Intensity Quantification}
Upon detection, storm intensity is quantified through visibility estimation:

\begin{equation}
V = \frac{566 \cdot r}{A_{\text{dust}} \cdot \lambda} \cdot \frac{\varepsilon''}{[(\varepsilon' + 2)^2 + \varepsilon''^2]}
\label{eq:visibility}
\end{equation}

where $A_{\text{dust}}$ is isolated from total attenuation using:

\begin{equation}
A_{\text{dust}} = A_{\text{total}} - A_0 - A_{\text{abs}} - A_{\text{fs}}
\label{eq:dust_isolation}
\end{equation}

\subsubsection{Error Analysis}

Given the Martian dust properties \cite{goldhirsh1982parameter}:

\begin{equation}
\bar{r} = 4 \times 10^{-6} \text{ m}, \quad
\varepsilon' = 1.55, \quad
\varepsilon'' = 6.3
\label{eq:dust_properties}
\end{equation}

Estimated uncertainties:
\begin{align}
\sigma_r &= 0.4 \times 10^{-6} \text{ m} \quad (10\%) \nonumber \\
\sigma_{\varepsilon'} &= 0.0775, \quad \sigma_{\varepsilon''} = 0.315 \quad (5\%) \nonumber \\
\sigma_N &= 0.2N \quad (20\%)
\label{eq:uncertainties}
\end{align}

The total attenuation is given by:
\begin{equation}
A_{\text{total}} = k(f) \cdot 10^4 \log_{10}{e} + \frac{1.029 \times 10^6 \cdot \varepsilon'' \cdot N \cdot r^3}{\left[(\varepsilon'+2)^2+\varepsilon''^2\right] \cdot \lambda}
\label{eq:total_attenuation}
\end{equation}

Error contributions from each variable:
\begin{align}
\sigma_{A_r}^2 &= \left(\frac{3.087 \times 10^6 \cdot \varepsilon'' \cdot N \cdot r^2}{\left[(\varepsilon'+2)^2+\varepsilon''^2\right] \cdot \lambda}\right)^2 \sigma_r^2 \label{eq:error_r} \\
\sigma_{A_N}^2 &= \left(\frac{1.029 \times 10^6 \cdot \varepsilon'' \cdot r^3}{\left[(\varepsilon'+2)^2+\varepsilon''^2\right] \cdot \lambda}\right)^2 (0.2N)^2 \label{eq:error_N} \\
\sigma_{A_{\varepsilon'}}^2 &= \left(\frac{4.116 \times 10^6 \cdot \varepsilon'' \cdot N \cdot r^3 \cdot (\varepsilon'+2)}{\left[(\varepsilon'+2)^2+\varepsilon''^2\right]^2 \cdot \lambda}\right)^2 \sigma_{\varepsilon'}^2 \label{eq:error_epsilon_real} \\
\sigma_{A_{\varepsilon''}}^2 &= \left(\frac{\partial A_{\text{total}}}{\partial \varepsilon''}\right)^2 \sigma_{\varepsilon''}^2 \label{eq:error_epsilon_imag}
\end{align}
where
\begin{equation}
\frac{\partial A_{\text{total}}}{\partial \varepsilon''} = \frac{1.029 \times 10^6 \cdot N \cdot r^3}{\left[(\varepsilon'+2)^2+\varepsilon''^2\right] \cdot \lambda} \left(1 - \frac{2\varepsilon''^2}{(\varepsilon'+2)^2+\varepsilon''^2}\right)
\label{eq:partial_epsilon_imag}
\end{equation}

The total variance in $A_{\text{total}}$ is:
\begin{equation}
\sigma_{A_{\text{total}}}^2 = \sigma_{A_r}^2 + \sigma_{A_N}^2 + \sigma_{A_{\varepsilon'}}^2 + \sigma_{A_{\varepsilon''}}^2
\label{eq:total_variance}
\end{equation}

For example, the signal of $f = 1$ THz ($\lambda = 3 \times 10^{-4}$ m) and $N = 10^8$ m$^{-3}$:
\begin{align*}
\sigma_{A_{\text{total},r}}^2 &= 1.97 \times 10^{-1} \text{ (dB/km)}^2 \\
\sigma_{A_{\text{total},\varepsilon'}}^2 &= 2.72 \times 10^{-3} \text{ (dB/km)}^2 \\
\sigma_{A_{\text{total},\varepsilon''}}^2 &= 4.51 \times 10^{-2} \text{ (dB/km)}^2
\end{align*}

The total standard deviation is thus:
\begin{equation}
\sigma_{A_{\text{total}}}  \approx 1.41 \text{ dB/km}
\label{eq:numerical_example}
\end{equation}

\begin{figure}[h!]
\centering
\includegraphics[width=0.45\textwidth]{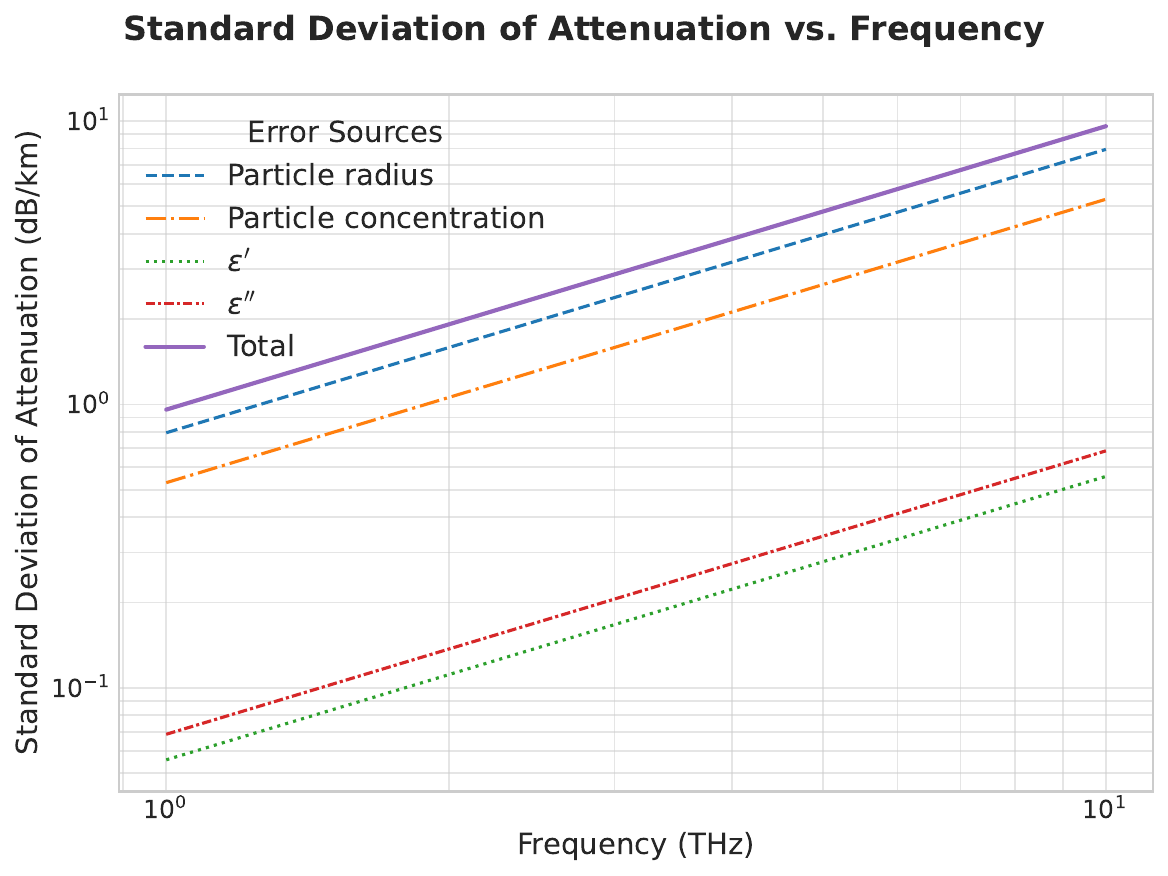}
\caption{Standard Deviation of Attenuation vs. Frequency for Martian Dust Storms. The plot shows the contribution of various error sources, including particle radius, particle concentration, and dielectric constants, to the overall uncertainty in signal attenuation across different THz frequencies.}
\label{fig:attenuation_uncertainty}
\end{figure}

The atmospheric attenuation characteristics under varying conditions, shown in Fig. \ref{fig:martian_atmosphere_attenuation}, demonstrate the system's sensitivity to environmental parameters. Additionally, Fig. \ref{fig:attenuation_uncertainty} illustrates the frequency-dependent uncertainty contributions from different error sources, providing crucial guidance for system optimization.

The analysis reveals frequency-dependent performance trade-offs and identifies particle characteristics as the dominant error source (Fig. \ref{fig:attenuation_uncertainty}).

\section{Performance Evaluation Metrics and Data Sources}

To assess the performance of the proposed MDSD system, we employ a set of evaluation metrics and utilize an authoritative Martian dust climatology database. This approach enables a robust comparison between the dust storm intensities detected by our THz communication link-based system and established observational data.

\subsection{Performance Metrics}

The MDSD system performance is evaluated using four complementary metrics: Mean Absolute Error (MAE), Correlation Coefficient ($\rho$), Normalized Bias (NBias), and Coverage. The MAE quantifies prediction accuracy through

\begin{equation}
\mathrm{MAE}=\frac{1}{T} \sum_{t=1}^{T}\left(\frac{1}{n} \sum_{j=1}^{n}\left|r_{jt}-r_{jt}^{\prime}\right|\right)
\end{equation}

where $r_{jt}^{\prime}$ and $r_{jt}$ represent predicted and true intensities for pixel $j$ at time $t$, with $T$ time steps and $n$ spatial pixels.

The correlation coefficient $\rho$ measures prediction-reference alignment:

% \begin{equation}
% \rho=\frac{1}{T} \sum_{t=1}^{T}\left(\frac{\sum_{j=1}^{n}\left(r_{jt}-\bar{r}{t}\right)\left(r{jt}^{\prime}-\bar{r}{t}^{\prime}\right)}{\sqrt{\sum{j=1}^{n}\left(r_{jt}-\bar{r}{t}\right)^{2}} \sqrt{\sum{j=1}^{n}\left(r_{jt}^{\prime}-\bar{r}_{t}^{\prime}\right)^{2}}}\right)
% \end{equation}

\begin{equation}
\rho=\frac{1}{T} \sum_{t=1}^{T}\left(\frac{\sum_{j=1}^{n}\left(r_{j t}-\bar{r}_t\right)\left(r_{jt}^{\prime}-\bar{r}_t^{\prime}\right)}
{\sqrt{\sum_{j=1}^{n}\left(r_{j t}-\bar{r}_t\right)^{2}} \sqrt{\sum_{j=1}^{n}\left(r_{jt}^{\prime}-\bar{r}_t^{\prime}\right)^{2}}}\right)
\end{equation}

where $\bar{r}_t$ and $\bar{r}_t^{\prime}$ denote mean true and predicted intensities at time $t$.

Systematic estimation bias is assessed through NBias:

\begin{equation}
\mathrm{NBias}=\frac{1}{T} \sum_{t=1}^{T}\left[\frac{\operatorname{Bias}(t)}{\left(\frac{1}{n} \sum_{j=1}^{n} r_{jt}^{\prime}\right)}\right]
\end{equation}

with $\operatorname{Bias}(t)=\frac{1}{n} \sum_{j=1}^{n}\left(r_{jt}-r_{jt}^{\prime}\right)$.

Finally, spatial completeness is quantified by Coverage:

\begin{equation}
\mathrm{Coverage} = \frac{\text{Number of valid pixels}}{\text{Total number of pixels}} \times 100\%
\end{equation}

This metric set enables comprehensive evaluation of the MDSD system's accuracy, tracking capability, systematic errors, and spatial completeness.

\subsection{Node Density Factor (NDF)}

To evaluate the MDSD system's performance across varying deployments, we introduce the NDF. This dimensionless metric characterizes the spatial distribution of communication nodes relative to the coverage area and maximum link length:
\begin{equation}
\mathrm{NDF} = \frac{N \cdot L_{\max}^2}{A}
\end{equation}
where:
\begin{itemize}
\item $N$ is the total number of communication nodes,
\item $L_{\max}$ is the maximum allowable link length (in km),
\item $A$ is the total coverage area (in km²).
\end{itemize}

This dimensionless metric provides crucial insights into:
\begin{itemize}
\item Optimal node distribution for Mars surface coverage
\item Trade-off between deployment cost and detection accuracy
\item Scalability of the MDSD system
\end{itemize}
The NDF analysis enables practical deployment planning by relating network density to detection performance under various Martian conditions.

\subsection{Martian Dust Climatology Database}

To evaluate the MDSD system, we utilize the Martian atmospheric dust climatology database provided by the Laboratoire de Météorologie Dynamique (LMD)~\cite{montabone2015eight}. This database offers a multi-annual climatology of airborne dust on Mars, spanning Martian years 24 to 36 (April 1999 to December 2022).

% The database provides gridded maps of column dust optical depth (CDOD) at 9.3~$\mu$m, derived from observations by multiple Mars orbiters including Mars Global Surveyor, Mars Odyssey, and Mars Reconnaissance Orbiter. The CDOD values are normalized to a reference pressure level of 610~Pa.

To relate the CDOD values from the database to the dust particle concentration $N$ in our model, we employ the following relationship:
\begin{equation}
N = \frac{\text{CDOD} \times 1.3}{Q_{\text{ext}} \pi r^2 H}
\end{equation}
where:
\begin{itemize}
    \item CDOD is the column dust optical depth from the database,
    \item The factor 1.3 converts absorption optical depth to extinction optical depth~\cite{lemmon2019martian},
    \item $Q_{\text{ext}}$ is the extinction efficiency factor, calculated as 3.57 for Martian dust at 9.3~$\mu$m using Mie theory,
    \item $r$ is the average dust particle radius, taken as 4~$\mu$m,
    \item $H$ is the scale height of the Martian atmosphere, approximately 11.1~km.
\end{itemize}

This relationship allows us to convert the CDOD values from the climatology database into dust particle concentrations, providing a direct comparison metric for evaluating the MDSD system's performance.

\section{Simulation and Performance Results}

To evaluate the effectiveness of the proposed MDSD system, we conducted extensive simulations using realistic Martian atmospheric data and various network configurations. Our simulations aim to assess the system's ability to accurately estimate dust storm intensities under different seasonal conditions and network densities.

\subsection{Simulation Setup}
The simulation environment was carefully designed to replicate the Martian atmosphere and dust storm conditions. We utilized high-fidelity data from the Mars Climate Database~\cite{montabone2015eight}, which provides information on Martian atmospheric conditions, including dust optical depth.

Table \ref{tab:sim_params} summarizes the key parameters used in our simulation setup.

\begin{table}[ht]
\centering
\caption{Simulation Parameters}
\label{tab:sim_params}
\begin{tabularx}{\linewidth}{|l|X|}
\hline
\textbf{Parameter} & \textbf{Value} \\
\hline
THz Frequency & 1 THz \\
\hline
Dust Particle Radius & 4 $\mu$m \\
\hline
Dust Dielectric Constant (Real Part) & 1.55 \\
\hline
Dust Dielectric Constant (Imaginary Part) & 6.3 \\
\hline
Maximum Link Length & 15 units \\
\hline
Number of Nodes & 20, 30, 50, 70, 100, 150, 200 \\
\hline
Simulation Duration & 10 Martian days per season \\
\hline
Seasons Analyzed & Dust storm (L$_s$ \(\approx\) 180°), Non-dust storm (L$_s$ \(\approx\) 0°) \\
\hline
Interpolation Methods & Linear, Nearest, Cubic, RBF, IDW, Kriging \\
\hline
\end{tabularx}
\end{table}

The simulation was conducted for two distinct Martian seasons: the dust storm season (around L$_s$ = 180°) and the non-dust storm season (around L$_s$ = 0°). For each season, we simulated 10 Martian days to capture the variability in atmospheric conditions.

We employed a range of NDFS to investigate the impact of network configuration on system performance. To estimate dust concentrations in areas without direct measurements, we implemented and compared six different interpolation methods: Linear, Nearest Neighbor, Cubic, Radial Basis Function (RBF), Inverse Distance Weighting (IDW), and Kriging.

The system's performance was evaluated using four metrics: MAE for error magnitude, Pearson for linear correlation, NBias for systematic bias, and Coverage for the extent of valid estimates, all compared to true values from the Mars Climate Database.

The simulation results provide valuable insights into the MDSD system's operation under various Martian conditions. 

\begin{figure*}[ht]
\centering
\subfloat[ Dust storm season]{
    \includegraphics[width=0.45\linewidth]{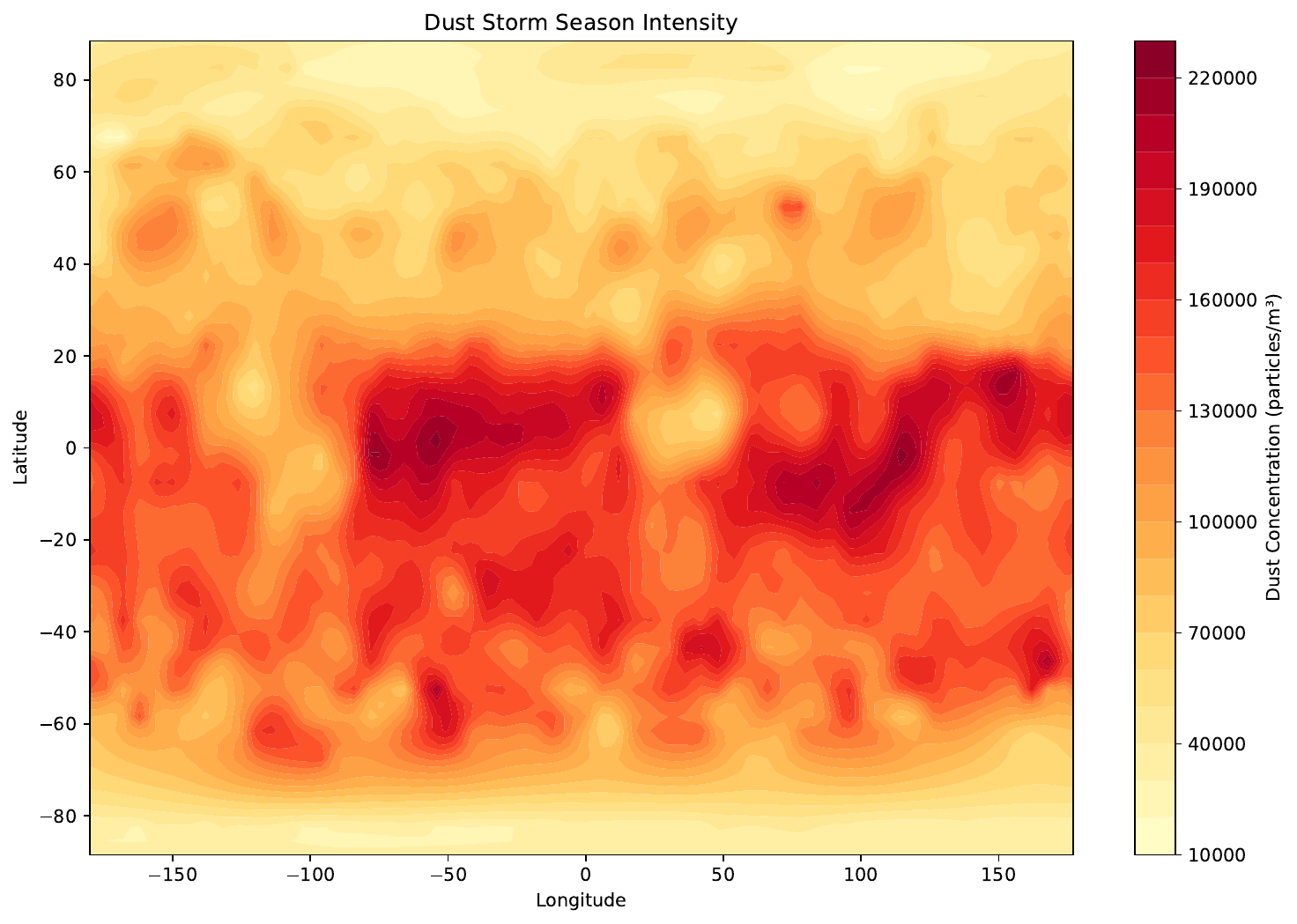}
    \label{fig:dust_intensity_a}
}
\hfill
\subfloat[ Non-dust storm season]{
    \includegraphics[width=0.45\linewidth]{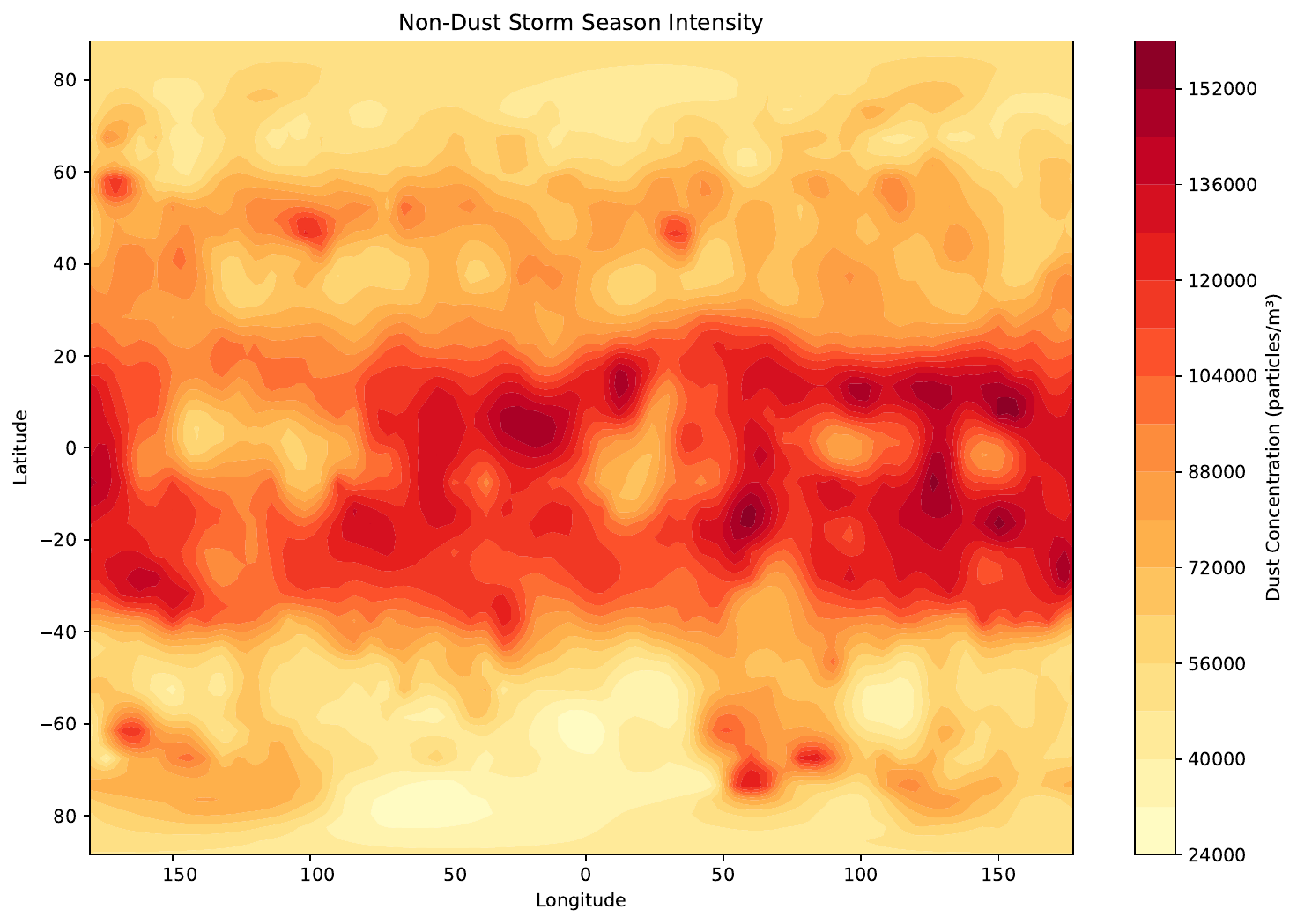}
    \label{fig:dust_intensity_b}
}
\caption{Simulated dust storm intensity on Mars}
\label{fig:dust_intensity}
\end{figure*}

Fig. \ref{fig:dust_intensity} shows the simulated dust storm intensity during different seasons. Fig. \ref{fig:dust_intensity_a} clearly shows regions of high dust concentration during the storm season, particularly in the southern hemisphere, while Fig. \ref{fig:dust_intensity_b} depicts relatively calm conditions with lower dust concentrations during the non-storm season. These visualizations derived from the Mars Climate Database serve as ground truth for our simulations.

\begin{figure*}[ht]
\centering
\subfloat[NDF = 1.67]{
    \includegraphics[width=0.3\textwidth]{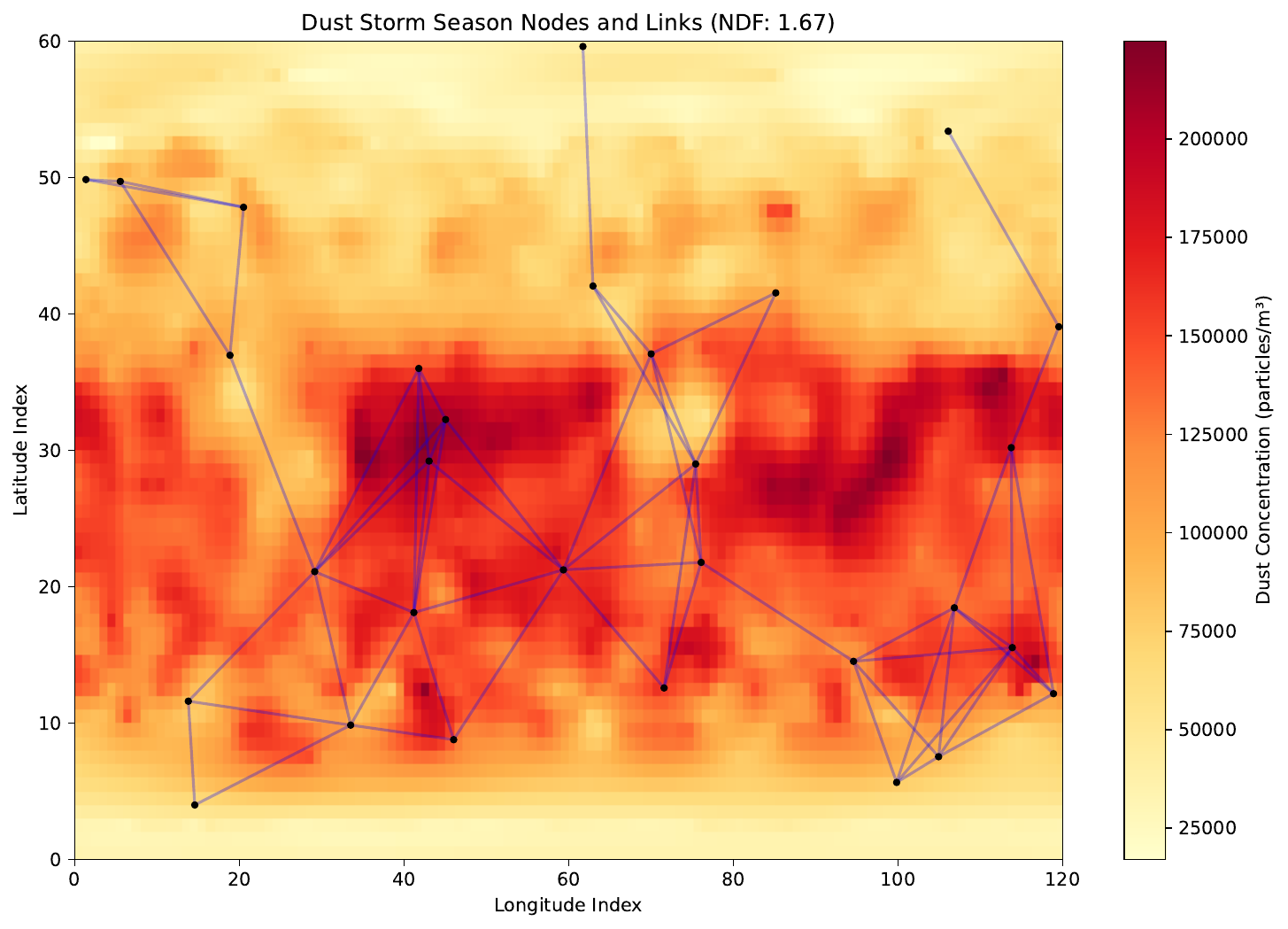}
    \label{fig:node_dist_a}
}
\hfill
\subfloat[NDF = 3.89]{
    \includegraphics[width=0.3\textwidth]{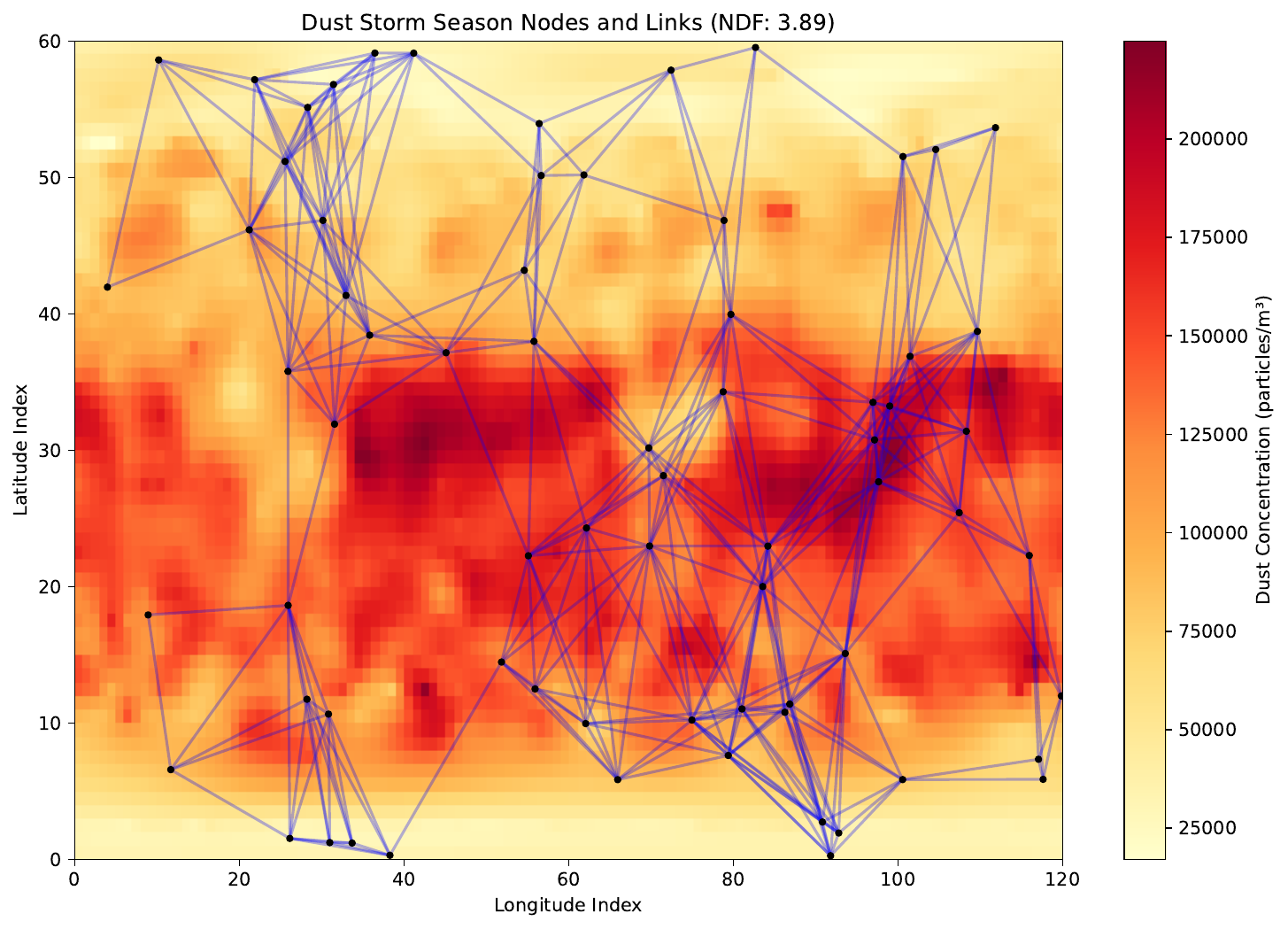}
    \label{fig:node_dist_b}
}
\hfill
\subfloat[NDF = 8.33]{
    \includegraphics[width=0.3\textwidth]{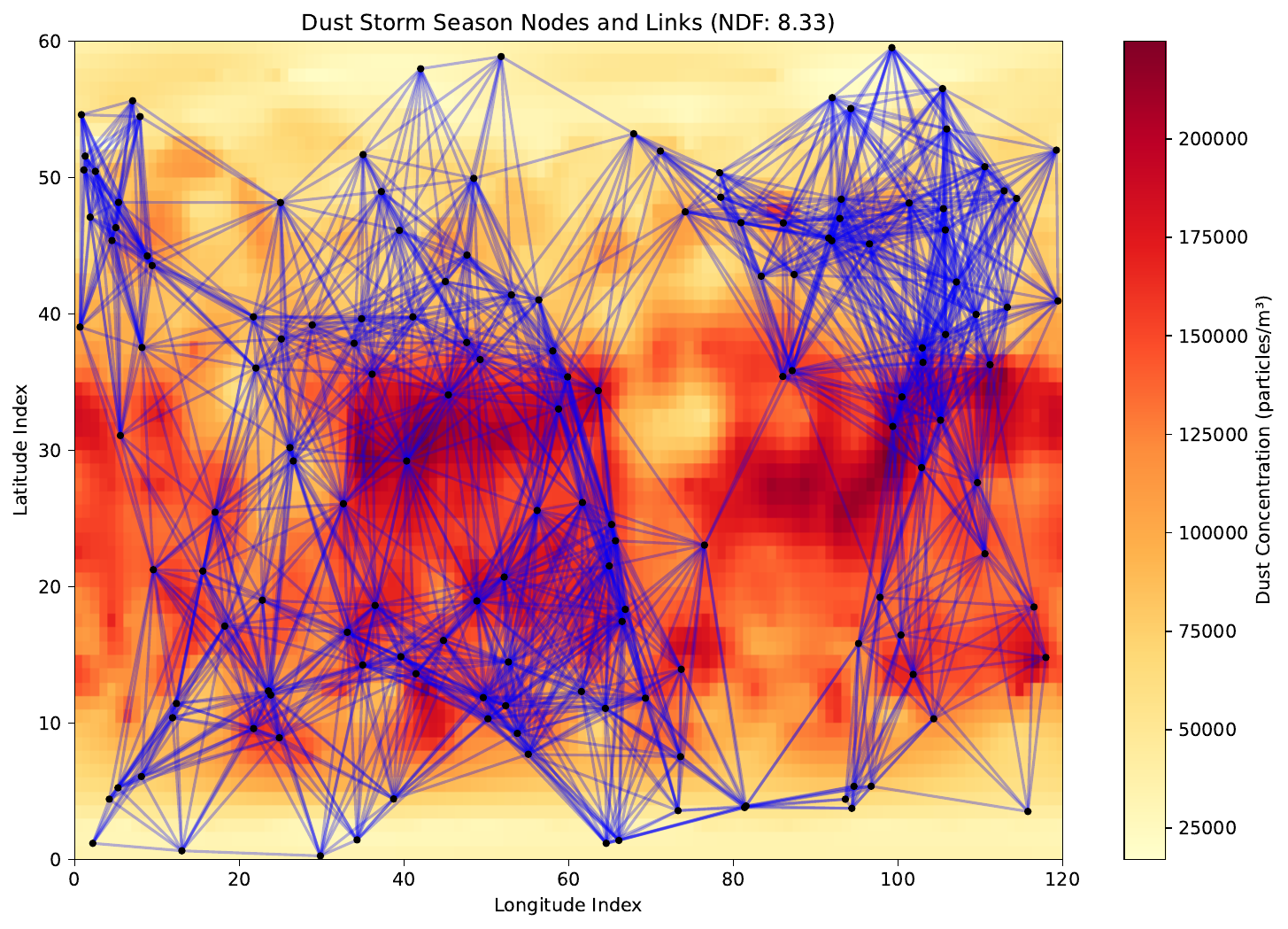}
    \label{fig:node_dist_c}
}
\caption{Node distribution and link configuration for various Node Density Factors (NDF)}
\label{fig:node_distribution}
\end{figure*}

To evaluate the impact of network density, we analyzed three scenarios with increasing Node Density Factors (NDFs) as shown in Fig. \ref{fig:node_distribution}. The sparse configuration (NDF=1.67) in Fig. \ref{fig:node_dist_a} represents limited infrastructure deployment. As the NDF increases to 3.89 and 8.33 in Fig. \ref{fig:node_dist_b} and Fig. \ref{fig:node_dist_c} respectively, we observe progressively denser networks reflecting potential future expansion of Martian communication assets. The background heatmap shows dust storm intensity, while black dots and blue lines represent node locations and communication links.

This visualization demonstrates how increasing network density provides more comprehensive surface coverage, potentially improving the accuracy and resolution of dust storm detection. The varying link lengths reflect the maximum 15-unit parameter used in our simulations, balancing long-distance communication needs with Martian atmospheric constraints. These results provide the foundation for subsequent performance analysis across different seasonal conditions and network configurations, offering crucial insights for optimizing the MDSD system's design and deployment strategies.

% The simulation results provide valuable insights into the MDSD system's operation and the challenges of dust storm detection on Mars. We present a series of visualizations that illustrate the simulated dust storm intensities, node distributions, and link configurations under various scenarios.

% \subsubsection{Dust Storm Intensity Visualization}

% Fig. \ref{fig:dust_intensity} shows the simulated dust storm intensity during the dust storm season (a) and non-dust storm season (b) on Mars. These visualizations are derived from the Mars Climate Database and serve as the ground truth for our simulations.

% Fig. \ref{fig:dust_intensity_a} clearly shows regions of high dust concentration during the storm season, particularly in the southern hemisphere. In contrast, Fig. \ref{fig:dust_intensity_b} depicts a relatively calm atmosphere with lower dust concentrations during the non-storm season. 

% \subsubsection{Node Distribution and Link Configuration}

% To illustrate the impact of network density on the MDSD system's performance, we present three scenarios with increasing NDFs in Fig. \ref{fig:node_distribution}.

% Fig. \ref{fig:node_dist_a} shows a sparse network configuration with an NDF of 1.67, representing a scenario with limited communication infrastructure on Mars. As the NDF increases to 3.89 in Fig. \ref{fig:node_dist_b} and 8.33 in Fig. \ref{fig:node_dist_c}, we observe a denser network with more nodes and links. These configurations reflect potential future expansion of Martian communication networks and allow us to assess the MDSD system's scalability.

% The background heatmap in each subfigure shows the dust storm intensity, whereas the black dots denote node locations and blue lines represent communication links. This illustration indicates how a denser network provides more thorough coverage of the Martian surface, which may enhance the accuracy and resolution of dust storm detection. These simulation outcomes form the basis of our following performance analysis. By adjusting the network density and testing the system under various seasonal conditions, we can evaluate the MDSD's resilience and efficiency in identifying and analyzing Martian dust storms across different scenarios. The knowledge gained from these simulations is essential for optimizing the design and deployment strategies of the MDSD system for upcoming Mars exploration missions.

\subsection{MAE Analysis of Interpolation Algorithms for MDSD}
\begin{figure}[ht]
\centering
\includegraphics[width=\linewidth]{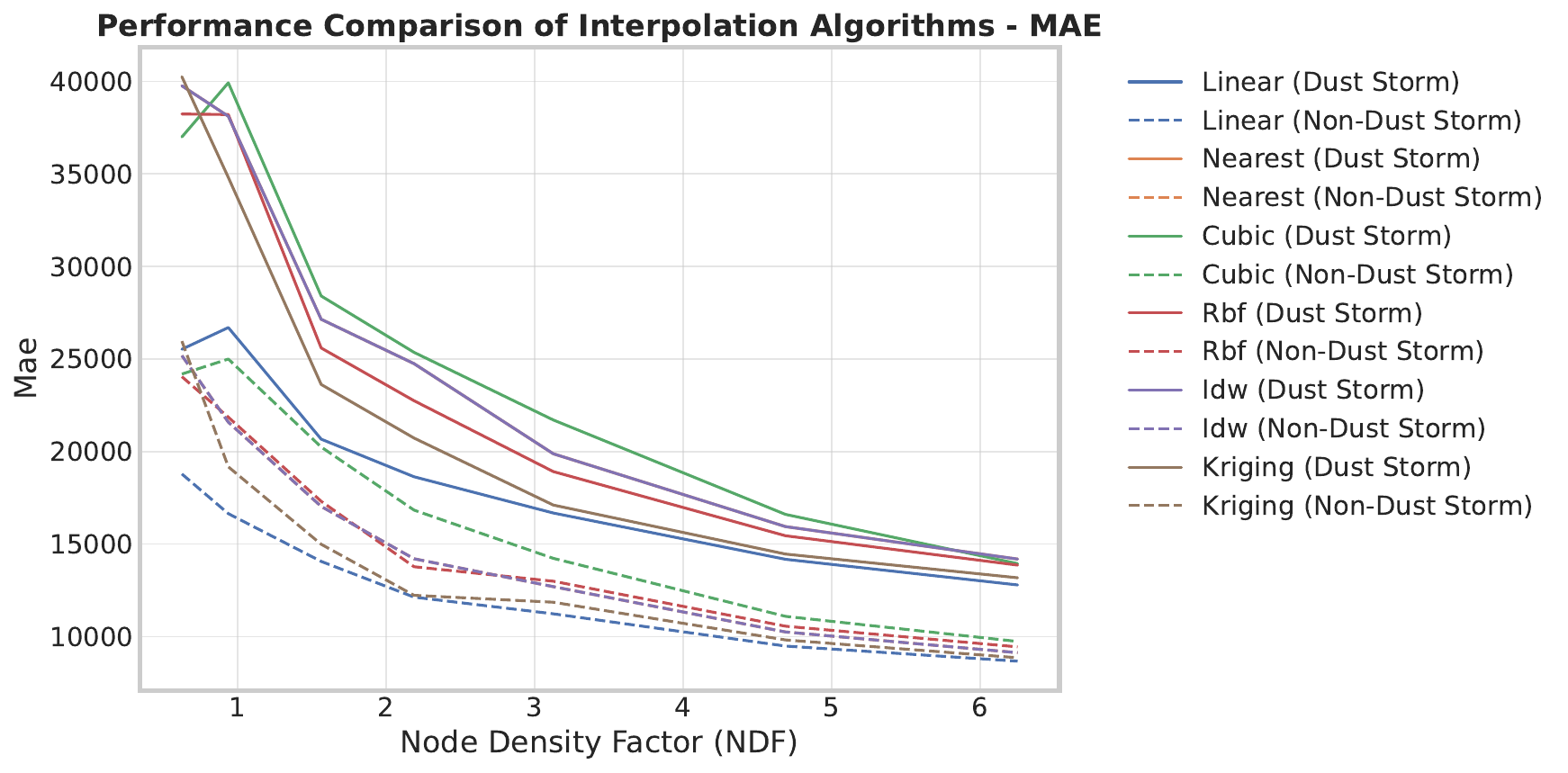}
\caption{MAE Performance of Interpolation Algorithms for Dust Storm Detection}
\label{fig:mae_performance}
\end{figure}

The Mean Absolute Error (MAE) performance analysis reveals distinct patterns across different interpolation algorithms (Fig. \ref{fig:mae_performance}). Linear interpolation consistently demonstrates superior performance, with MAE values approximately 50\% lower than other methods across all NDFs in both storm and non-storm conditions. Kriging shows comparable performance at higher NDFs, particularly in non-storm scenarios, while nearest neighbor and IDW algorithms exhibit identical performance patterns, suggesting similar interpolation approaches in the Martian environment.

The cubic interpolation method shows the highest MAE values, averaging 30\% higher than linear interpolation during dust storms, indicating its limited suitability for this application. RBF performance falls between the linear and cubic methods, offering moderate accuracy across all conditions.

All algorithms show significant improvement with increasing NDF, with average MAE reductions of 40-60\% between lowest and highest NDFs. This trend strongly suggests that denser ISAC device deployment could substantially enhance the MDSD system's accuracy, especially during dust storm events.

\subsection{Correlation Analysis of Interpolation Algorithms}
\begin{figure}[ht]
\centering
\includegraphics[width=\linewidth]{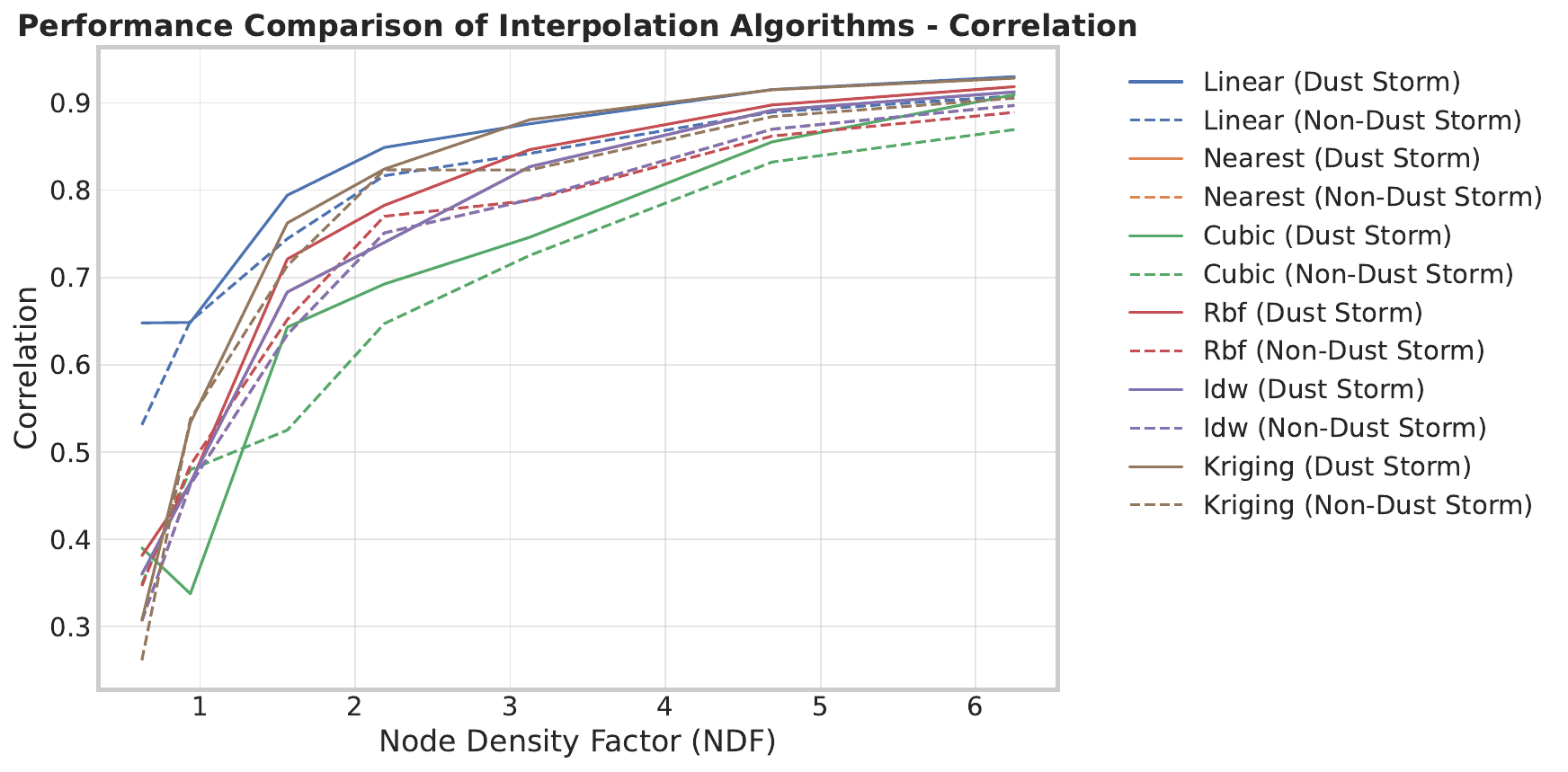}
\caption{Correlation Performance of Interpolation Algorithms for Dust Storm Detection}
\label{fig:correlation_performance}
\end{figure}

The correlation analysis reveals distinct performance patterns among interpolation algorithms under varying network densities (Fig. \ref{fig:correlation_performance}). Linear interpolation achieves the highest correlation coefficients, ranging from 0.65 to 0.93 during dust storms and maintaining robust performance (0.53-0.91) in non-storm conditions. Kriging demonstrates comparable effectiveness at high NDFs, particularly during non-storm periods where it approaches linear interpolation's performance (correlation $>$ 0.90).

Nearest neighbor and IDW algorithms show identical moderate performance patterns, suggesting similar interpolation mechanisms in the Martian environment. Cubic interpolation exhibits the most pronounced NDF dependency, with correlation values improving from 0.34 at low NDFs to approximately 0.90 at maximum density, indicating its potential utility only in dense network deployments.

The consistent improvement in correlation coefficients with increasing NDF across all algorithms (average increase of 0.3-0.4) underscores the significant advantage of denser ISAC device deployment for accurate dust storm monitoring.

\subsection{Normalized Bias Analysis}
\begin{figure}[ht]
\centering
\includegraphics[width=\linewidth]{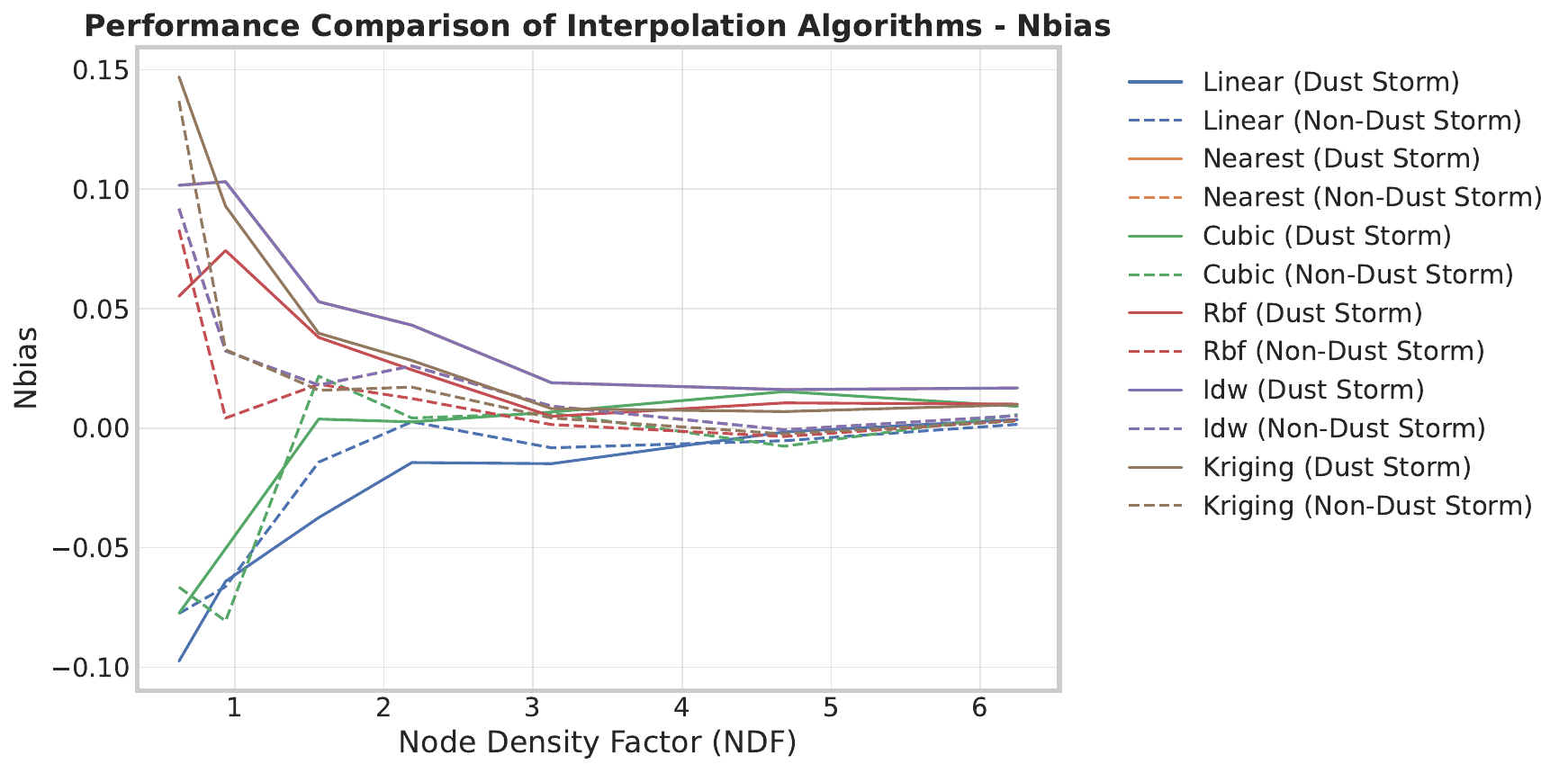}
\caption{Normalized Bias Performance of Interpolation Algorithms for Dust Storm Detection}
\label{fig:nbias_performance}
\end{figure}

Analysis of Normalized Bias (NBias) reveals systematic estimation tendencies across interpolation algorithms (Fig. \ref{fig:nbias_performance}). At low NDFs, linear interpolation exhibits a negative bias (-0.097), while kriging shows the highest positive bias (0.147), representing contrasting approaches to dust storm intensity estimation. The conservative nature of linear interpolation may offer advantages in applications where false alarm minimization is critical.

Algorithm performance converges significantly as network density increases. Linear interpolation achieves near-zero bias (±0.004) at maximum NDF, while kriging and RBF methods demonstrate similar convergence (±0.01). Cubic interpolation shows the most variable performance, transitioning between negative and positive bias values across the NDF range, suggesting limited reliability in sparse network configurations.

The systematic reduction in bias magnitude with increasing NDF (average reduction of 85\% across all algorithms) quantitatively demonstrates the importance of network density optimization in the MDSD system design. This trend, consistent across all interpolation methods, provides clear guidance for future Martian network deployment strategies.

\subsection{Coverage Analysis}
\begin{figure}[ht]
\centering
\includegraphics[width=\linewidth]{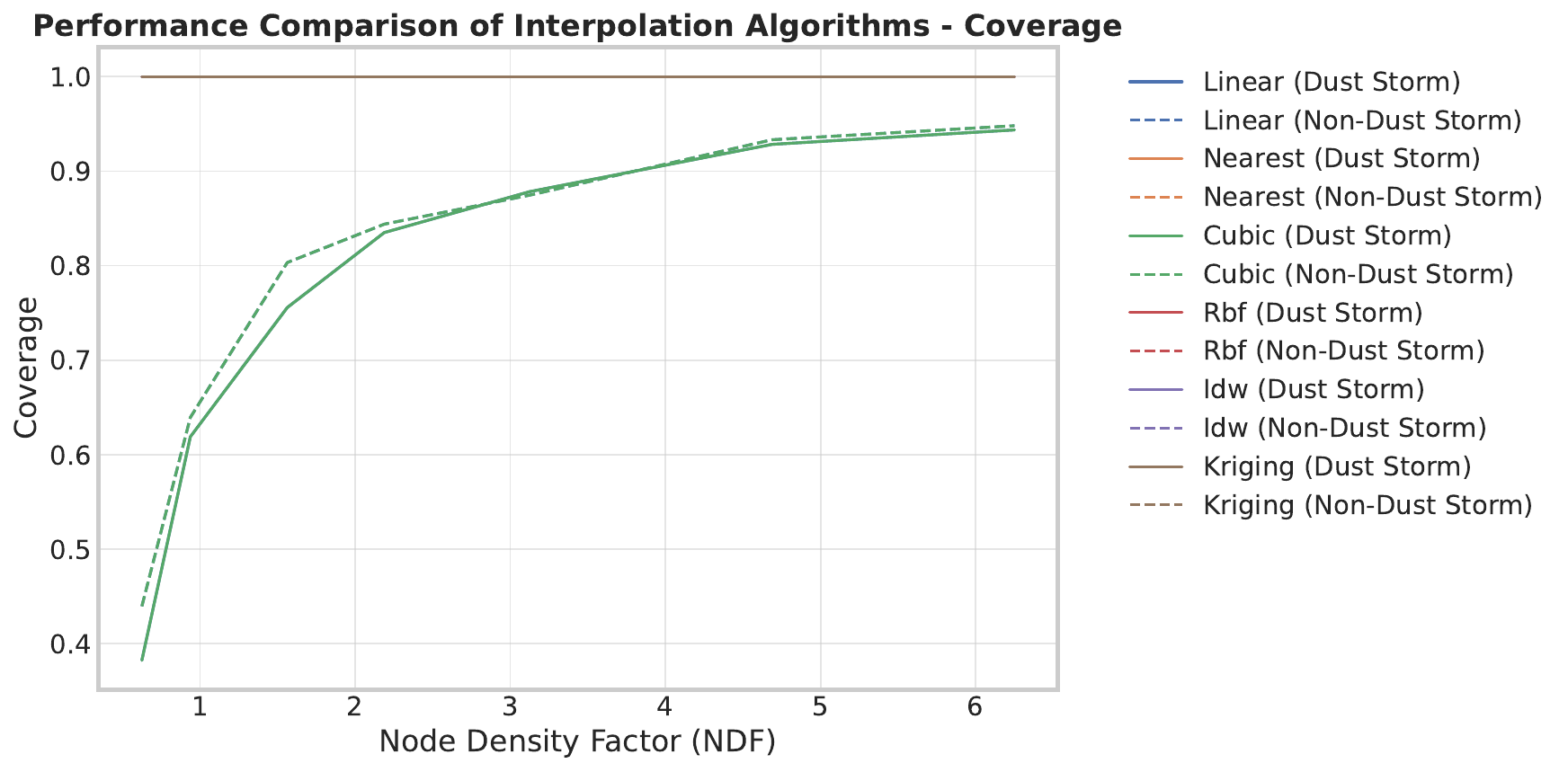}
\caption{Coverage Performance of Interpolation Algorithms for Dust Storm Detection}
\label{fig:coverage_performance}
\end{figure}

Coverage analysis reveals a clear algorithmic dichotomy (Fig. \ref{fig:coverage_performance}). The first group - nearest neighbor, RBF, IDW, and kriging algorithms - maintains 100\% coverage across all NDFs, providing comprehensive monitoring capability even in sparse networks. The second group - linear and cubic interpolation methods - demonstrates NDF-dependent coverage, improving from 40\% at NDF=0.625 to 95\% at NDF=6.25, with the most significant enhancement occurring between NDFs of 0.625 and 2.1875.

This performance division suggests an optimal hybrid approach for the MDSD system: employing full-coverage algorithms (e.g., kriging) for broad-scale monitoring, supplemented by linear interpolation in high-density regions where accuracy is paramount. Such an adaptive strategy effectively balances coverage requirements with precision needs, particularly relevant for the varying network densities characteristic of Martian deployments.

\section{Conclusion}

This paper presented the Mars Dust Storm Detector (MDSD), a THz ISAC-based system for Martian dust storm monitoring. Through comprehensive attenuation modeling, performance analysis of interpolation algorithms and error characterization, we demonstrated the system's capability to provide reliable dust storm detection using future communication infrastructure.

Performance evaluation revealed that linear interpolation achieves superior accuracy (correlation $>$ 0.90 at high NDFs), while full-coverage algorithms maintain complete spatial monitoring in sparse networks. Error analysis identified dust particle size uncertainty as the primary contributor to estimation errors, though the system demonstrated resilience to Martian atmospheric variations.

Key findings include:
\begin{itemize}
\item Network density significantly impacts detection accuracy, with performance improvements of up to 85\% at maximum NDF
\item Adaptive interpolation strategy optimizes the coverage-accuracy trade-off
\item System robustness across typical Martian environmental variations
\end{itemize}

Future work should focus on system validation through Earth-based analogs and integration with Mars networks. The MDSD system demonstrates the potential of ISAC technology for efficient environmental monitoring in planetary exploration, providing a foundation for enhanced scientific understanding and future Mars mission planning.

% \correspauthor%

\renewcommand{\refname}{\small\normalfont\MakeUppercase{References}} % Adjust font size as needed
\bibliographystyle{ieeetr} % Choose IEEE style
\bibliography{references}  % Use your BibTeX file (without .bib extension)

\begin{thebibliography}{10}

\bibitem{IoSp}
I.~Akyildiz, O.~Akan, C.~Chen, J.~Fang, and W.~Su, ``The state of the art in interplanetary internet,'' {\em IEEE Communications Magazine}, vol.~42, no.~7, pp.~108--118, 2004.

\bibitem{wedage2022path}
L.~T. Wedage, B.~Butler, S.~Balasubramaniam, M.~C. Vuran, and Y.~Koucheryavy, ``Path loss analysis of terahertz communication in mars' atmospheric conditions,'' in {\em 2022 IEEE International Conference on Communications Workshops (ICC Workshops)}, pp.~1225--1230, IEEE, 2022.

\bibitem{goldstein1968communication}
B.~S. Goldstein, ``Communication from mars: Requirements and limitations,'' {\em IEEE Transactions on Aerospace and Electronic Systems}, no.~3, pp.~392--401, 1968.

\bibitem{doexamining}
K.~Do and R.~Yelle, ``Examining the 2018 summer global dust storm on mars’s lower atmosphere from nasa’s mars reconnaissance orbiter,''

\bibitem{balme2006dust}
M.~Balme and R.~Greeley, ``Dust devils on earth and mars,'' {\em Reviews of Geophysics}, vol.~44, no.~3, 2006.

\bibitem{liu2022integrated}
F.~Liu, Y.~Cui, C.~Masouros, J.~Xu, T.~X. Han, Y.~C. Eldar, and S.~Buzzi, ``Integrated sensing and communications: Toward dual-functional wireless networks for 6g and beyond,'' {\em IEEE journal on selected areas in communications}, vol.~40, no.~6, pp.~1728--1767, 2022.

\bibitem{liu2022survey}
A.~Liu, Z.~Huang, M.~Li, Y.~Wan, W.~Li, T.~X. Han, C.~Liu, R.~Du, D.~K.~P. Tan, J.~Lu, {\em et~al.}, ``A survey on fundamental limits of integrated sensing and communication,'' {\em IEEE Communications Surveys \& Tutorials}, vol.~24, no.~2, pp.~994--1034, 2022.

\bibitem{pawar2013terahertz}
A.~Y. Pawar, D.~D. Sonawane, K.~B. Erande, and D.~V. Derle, ``Terahertz technology and its applications,'' {\em Drug invention today}, vol.~5, no.~2, pp.~157--163, 2013.

\bibitem{messer2006environmental}
H.~Messer, A.~Zinevich, and P.~Alpert, ``Environmental monitoring by wireless communication networks,'' {\em Science}, vol.~312, no.~5774, pp.~713--713, 2006.

\bibitem{messer2015new}
H.~Messer and O.~Sendik, ``A new approach to precipitation monitoring: A critical survey of existing technologies and challenges,'' {\em IEEE Signal Processing Magazine}, vol.~32, no.~3, pp.~110--122, 2015.

\bibitem{wedage2021climate}
L.~T. Wedage, B.~Butler, S.~Balasubramaniam, Y.~Koucheryavy, and J.~M. Jornet, ``Climate change sensing through terahertz communications: A disruptive application of 6g networks,'' {\em arXiv preprint arXiv:2110.03074}, 2021.

\bibitem{leijnse2007rainfall}
H.~Leijnse, R.~Uijlenhoet, and J.~Stricker, ``Rainfall measurement using radio links from cellular communication networks,'' {\em Water resources research}, vol.~43, no.~3, 2007.

\bibitem{david2009novel}
N.~David, P.~Alpert, and H.~Messer, ``Novel method for water vapour monitoring using wireless communication networks measurements,'' {\em Atmospheric chemistry and physics}, vol.~9, no.~7, pp.~2413--2418, 2009.

\bibitem{olsen1978ar}
R.~Olsen, D.~V. Rogers, and D.~Hodge, ``The ar b relation in the calculation of rain attenuation,'' {\em IEEE Transactions on antennas and propagation}, vol.~26, no.~2, pp.~318--329, 1978.

\bibitem{dong2024debrisense}
H.~Dong and O.~B. Akan, ``Debrisense: Thz-based integrated sensing and communications (isac) for debris detection and classification in the internet of space (ios),'' {\em IEEE Transactions on Wireless Communications}, 2024.
\newblock Under review.

\bibitem{diao2021comparison}
Z.~Diao, Q.~Jing, and W.~Zhong, ``Comparison of the influence of martian and earth's atmospheric environments on terahertz band electromagnetic waves,'' {\em International Journal of Communication Systems}, vol.~34, no.~12, p.~e4894, 2021.

\bibitem{wedage2023comparative}
L.~T. Wedage, B.~Butler, S.~Balasubramaniam, Y.~Koucheryavy, and M.~C. Vuran, ``Comparative analysis of terahertz propagation under dust storm conditions on mars and earth,'' {\em IEEE Journal of Selected Topics in Signal Processing}, 2023.

\bibitem{tekbiyik2022wireless}
K.~Tekb{\i}y{\i}k, D.~Altinel, M.~Cansiz, and G.~K. Kurt, ``Wireless power transmission on martian surface for zero-energy devices,'' {\em IEEE Transactions on Aerospace and Electronic Systems}, vol.~58, no.~5, pp.~3870--3880, 2022.

\bibitem{wang2015origin}
H.~Wang and M.~I. Richardson, ``The origin, evolution, and trajectory of large dust storms on mars during mars years 24--30 (1999--2011),'' {\em Icarus}, vol.~251, pp.~112--127, 2015.

\bibitem{shekh2021effect}
N.~A. Shekh, V.~Dviwedi, and J.~P. Pabari, ``Effect of sandstorm on radio propagation model of mars,'' in {\em International Conference on Mobile Computing and Sustainable Informatics: ICMCSI 2020}, pp.~441--447, Springer, 2021.

\bibitem{alozie2023review}
E.~Alozie, A.~Musa, N.~Faruk, A.~L. Imoize, A.~Abdulkarim, A.~D. Usman, Y.~O. Imam-Fulani, K.~S. Adewole, A.~A. Oloyede, O.~A. Sowande, {\em et~al.}, ``A review of dust-induced electromagnetic waves scattering theories and models for 5g and beyond wireless communication systems,'' {\em Scientific African}, p.~e01816, 2023.

\bibitem{goldhirsh2001attenuation}
J.~Goldhirsh, ``Attenuation and backscatter from a derived two-dimensional duststorm model,'' {\em IEEE Transactions on Antennas and Propagation}, vol.~49, no.~12, pp.~1703--1711, 2001.

\bibitem{hulst1981light}
H.~C. Hulst and H.~C. van~de Hulst, {\em Light scattering by small particles}.
\newblock Courier Corporation, 1981.

\bibitem{smith1986propagation}
E.~Smith and W.~Flock, ``Propagation through martian dust at 8.5 and 32 ghz,'' {\em The Telecommunications and Data Acquisition Report}, 1986.

\bibitem{o2019perspective}
J.~F. O’Hara, S.~Ekin, W.~Choi, and I.~Song, ``A perspective on terahertz next-generation wireless communications,'' {\em Technologies}, vol.~7, no.~2, p.~43, 2019.

\bibitem{smith2008martian}
M.~D. Smith, ``The martian atmosphere: A review of results from the mars global surveyor mission,'' {\em Annual Review of Earth and Planetary Sciences}, vol.~36, pp.~191--219, 2008.

\bibitem{5995306}
J.~M. Jornet and I.~F. Akyildiz, ``Channel modeling and capacity analysis for electromagnetic wireless nanonetworks in the terahertz band,'' {\em IEEE Transactions on Wireless Communications}, vol.~10, no.~10, pp.~3211--3221, 2011.

\bibitem{gordon2022hitran}
I.~E. Gordon, L.~S. Rothman, R.~J. Hargreaves, R.~Hashemi, E.~V. Karlovets, F.~M. Skinner, E.~K. Conway, C.~Hill, R.~V. Kochanov, Y.~Tan, {\em et~al.}, ``Hitran2020: part 1. line lists for h2o, co2, o3, n2o, co, ch4, nh3, hf, hcl, hbr, and h2s,'' {\em Journal of Quantitative Spectroscopy and Radiative Transfer}, vol.~277, p.~107949, 2022.

\bibitem{mahaffy2013abundance}
P.~R. Mahaffy, C.~R. Webster, S.~K. Atreya, H.~Franz, M.~Wong, P.~G. Conrad, D.~Harpold, J.~J. Jones, L.~A. Leshin, H.~Manning, {\em et~al.}, ``Abundance and isotopic composition of gases in the martian atmosphere from the curiosity rover,'' {\em Science}, vol.~341, no.~6143, pp.~263--266, 2013.

\bibitem{ho2002radio}
C.~Ho, N.~Golshan, and A.~Kliore, ``Radio wave propagation handbook for communication on and around mars,'' tech. rep., 2002.

\bibitem{sahin2024resource}
S.~Sahin and T.~Girici, ``Resource allocation in networked joint radar and communications,'' {\em IEEE Transactions on Aerospace and Electronic Systems}, 2024.

\bibitem{goldhirsh1982parameter}
J.~Goldhirsh, ``A parameter review and assessment of attenuation and backscatter properties associated with dust storms over desert regions in the frequency range of 1 to 10 ghz,'' {\em IEEE Transactions on Antennas and Propagation}, vol.~30, no.~6, pp.~1121--1127, 1982.

\bibitem{montabone2015eight}
L.~Montabone, F.~Forget, E.~Millour, R.~Wilson, S.~Lewis, B.~Cantor, D.~Kass, A.~Kleinb{\"o}hl, M.~Lemmon, M.~Smith, {\em et~al.}, ``Eight-year climatology of dust optical depth on mars,'' {\em Icarus}, vol.~251, pp.~65--95, 2015.

\bibitem{lemmon2019martian}
M.~Lemmon, S.~Guzewich, T.~McConnochie, G.~Mart{\'\i}nez, A.~de~Vicente-Retortillo, M.~Smith, J.~Bell, D.~Wellington, and S.~Jacobs, ``Martian dust particle size during the 2018 planet-encircling dust storm as measured by the curiosity rover,'' in {\em Ninth International Conference on Mars}, vol.~2089, p.~6298, 2019.

\end{thebibliography}

\begin{IEEEbiography}[\includegraphics{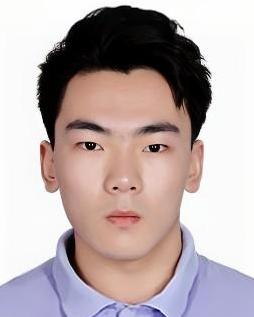}]{Haofan Dong}(Student Member, IEEE) is currently pursuing the Ph.D. degree with the Internet of Everything (IoE) Group, Department of Engineering, University of Cambridge, Cambridge, UK. He received the B.Eng. degree in from University of Leeds, Leeds, UK, in 2022, and the M.Res. degree in Connected Electronic and Photonic Systems from University College London, London, UK, in 2023.

His research interests include integrated sensing and communications (ISAC), Terahertz communications, and space communications.
\end{IEEEbiography}

\begin{IEEEbiography}[\includegraphics{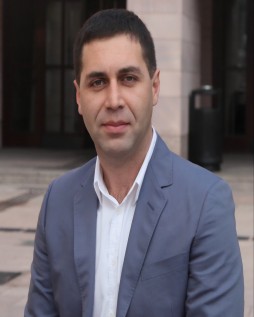}]{Ozgur B. Akan}(Fellow, IEEE) received the Ph.D. degree from the School of Electrical and Computer Engineering, Georgia Institute of Technology, Atlanta, in 2004. He is currently the Head of the Internet of Everything Group, Department of Engineering, University of Cambridge, U.K., and the Director of the Centre for next-generation Communications, Koç University, Türkiye. His research interests include wireless, nano, and molecular communications, and Internet of Everything.
\end{IEEEbiography}

\end{document}